\documentclass[Journal, 10pt]{IEEEtran}
\IEEEoverridecommandlockouts
\usepackage{cite}
\usepackage{amsmath,amssymb,amsfonts}
\usepackage{graphicx}
\usepackage{textcomp}

\usepackage{bbm}

\usepackage{xcolor}
\usepackage{color,soul}
\usepackage{epstopdf}
\usepackage{subcaption}

\usepackage{algorithm}
\usepackage[noend]{algpseudocode}

\usepackage{cases}

\def\BibTeX{{\rm B\kern-.05em{\sc i\kern-.025em b}\kern-.08em
                T\kern-.1667em\lower.7ex\hbox{E}\kern-.125emX}}
\newtheorem{definition}{Definition}

\newtheorem{lemma}{Lemma}

\def\bq{{\bf q}}
\def\bu{{\bf u}}

\def\bL{{\bf L}}
\def\bx{{\bf x}}
\def\by{{\bf y}}
\makeatletter
\newcommand*{\rom}[1]{\expandafter\@slowromancap\romannumeral #1@}
\renewcommand{\Function}[2]{%
        \csname ALG@cmd@\ALG@L @Function\endcsname{#1}{#2}%
        \def\jayden@currentfunction{#1}%
}
\newcommand{\funclabel}[1]{%
        \@bsphack
        \protected@write\@auxout{}{%
                \string\newlabel{#1}{{\jayden@currentfunction}{\thepage}}%
        }%
        \@esphack
}
\makeatother

\IEEEpubid{\begin{minipage}[t]{\textwidth}\ \\[10pt]
		\centering\scriptsize{\textcopyright~Copyright (c) 2015 IEEE. Personal use of this material is permitted. However, permission to use this material for any other purposes must be obtained from the IEEE by sending a request to pubs-permissions@ieee.org.}
\end{minipage}} 
\begin{document}
%\onecolumn
\title{Coarse Trajectory Design for Energy Minimization in UAV-enabled Wireless Communications with Latency Constraints}
\author{$\text{Dinh-Hieu Tran}^*, \textit{Student Member, IEEE}$, $\text{Thang X. Vu}^*, \textit{Member, IEEE}$, $\text{Symeon Chatzinotas}^*, \textit{Senior Member, IEEE},$  $\text{Shahram ShahbazPanahi}^{*,\dagger}, \textit{Senior Member, IEEE}$, $\text{and Bj{\"o}rn Ottersten}^*, \textit{Fellow, IEEE}$
	\\ ${^*}$Interdisciplinary Centre for Security, Reliability and Trust (SnT), the University of Luxembourg, Luxembourg.\\
       ${^\dagger}$Department of Electrical, Computer, and Software Engineering, University of Ontario Institute of Technology, Oshawa, ON L1H 7K4, Canada\\
         Email: \{hieu.tran-dinh, thang.vu, symeon.chatzinotas, bjorn.ottersten\}@uni.lu, shahram.shahbazpanahi@uoit.ca
%               {\footnotesize \textsuperscript{*}Note: Sub-titles are not captured in Xplore and should not be used}
       
        \thanks{The research leading to these results has received funding from the Luxembourg National Research Fund under project FNR CORE ProCAST, grant C17/IS/11691338 and FNR 5G-Sky, grant C19/IS/13713801.}
         \thanks{ Part of this work has been presented in IEEE Asilomar Conference on Signals, Systems, and Computers, 3-6 November 2019, Pacific Grove, CA, USA \cite{Hieu}.}
        \thanks{The authors are with the Interdisciplinary Centre for Security, Reliability and Trust (SnT), University of Luxembourg, L-1855 Luxembourg. (e-mail: \{hieu.tran-dinh, thang.vu, symeon.chatzinotas, bjorn.ottersten\}@uni.lu, shahram.shahbazpanahi@uoit.ca).}
        \thanks{The authors would like to extend their sincere thanks to Ashok Bandi and Sumit Gautam for providing valuable feedback to help improve the quality of the manuscript.}
        \thanks{Corresponding author: Dinh-Hieu~Tran (e-mail: hieu.tran-dinh@uni.lu).}
        
}

\maketitle

\vspace{-5cm}

%%%%%%%%%%%%%%%%%%%%%%%%%%%%%%%%%%%%%%%%%%%%%%%%%%%%%%%%%%%%%%%%%%%%%%
%                              ABSTRACT                              %
%%%%%%%%%%%%%%%%%%%%%%%%%%%%%%%%%%%%%%%%%%%%%%%%%%%%%%%%%%%%%%%%%%%%%%
\begin{abstract}
In this paper, we design the UAV trajectory to minimize the total energy consumption while satisfying the requested timeout (RT) requirement and energy budget, which is accomplished via jointly optimizing the path and UAV's velocities along subsequent hops. The corresponding optimization problem is difficult to solve due to its non-convexity and combinatorial nature. To overcome this difficulty, we solve the original problem via two consecutive steps. Firstly, we propose two algorithms, namely heuristic search, and dynamic programming (DP) to obtain a feasible set of paths without violating the GU's RT requirements based on the traveling salesman problem with time window (TSPTW). Then, they are compared with exhaustive search and traveling salesman problem (TSP) used as reference methods. While the exhaustive algorithm achieves the best performance at a high computation cost, the heuristic algorithm exhibits poorer performance with low complexity. As a result, the DP is proposed as a practical trade-off between the exhaustive and heuristic algorithms. Specifically, the DP algorithm results in near-optimal performance at a much lower complexity. Secondly, for given feasible paths, we propose an energy minimization problem via a joint optimization of the UAV's velocities along subsequent hops. Finally, numerical results are presented to demonstrate the effectiveness of our proposed algorithms. The results show that the DP-based algorithm approaches the exhaustive search's performance with a significantly reduced complexity. It is also shown that the proposed solutions outperform the state-of-the-art benchmarks in terms of both energy consumption and outage performance.

\end{abstract}


%%%%%%%%%%%%%%%%%%%%%%%%%%%%%%%%%%%%%%%%%%%%%%%%%%%%%%%%%%%%%%%%%%%%%%
%                           Introduction                             %
%%%%%%%%%%%%%%%%%%%%%%%%%%%%%%%%%%%%%%%%%%%%%%%%%%%%%%%%%%%%%%%%%%%%%%
\section{Introduction}

With the proliferation of mobile devices and data-hungry applications, the next generation wireless networks are expected to support not only the unprecedented traffic increase and stringent latency but also ubiquitous coverage requirements. Although heterogeneous networks (HetNets) \cite{Hetnet} and cloud radio access networks (C-RANs) \cite{Vu,Vu_1} have shown their capability in supporting massive network traffics, their deployments are usually focused on dense areas. In less-dense areas, e.g., urban, and places where the network traffic highly fluctuates, the employment of C-RANs is economically inefficient. In such cases, the current terrestrial network architecture might suffer network congestion or be unable to support the ubiquitous coverage. 

Recently, unmanned aerial vehicles (UAVs) have attracted much attention as a promising solution for improving the performance of terrestrial wireless communication networks thanks to their mobility, agility, and flexible deployment \cite{Mozaffari}. By employing a flying base station, UAVs can be deployed along with ground base stations (GBSs) to provide pervasive coverage and timely applications to ground users (GUs). Consequently, the deployment of UAVs in wireless communications has found applications in various domains, such as disaster rescue mission \cite{Erdelj}, surveillance\cite{Li}, and smart farming \cite{Bacco}. Besides many advantages, UAV-enabled communications are not without limitation. The inherent limitations of UAVs has imposed technical restrictions on size, weight, and power capability (SWAP), which consequently affect the UAV's endurance and performance \cite{Y_Zeng}. One of the major challenges in UAV deployment is to efficiently design the trajectory in order to maximize the UAV's service lifetime.

Certain efforts have recently been devoted to efficient UAV trajectory design \cite{D_Yang,Y_Zeng,Phu,Cai,Hua,Sun, Guo, Kang}. Yang et al. in \cite{D_Yang} investigate the different Pareto efficiency between the optimal GU transmit power and UAV trajectory design. Phu et al. \cite{Phu} use UAV as a friendly jammer to maximize the average secrecy rate of the cognitive radio network (CRN) by jointly optimizing the transmission power and UAV trajectory. Reference \cite{Y_Zeng} designs the trajectory of UAV to minimize the mission completion time in UAV-enabled multi-casting systems based on the traveling salesman problem (TSP). References \cite{Cai,Hua,Xiang} study more complicated scenarios with multiple UAVs. The authors of \cite{Cai} investigate the dual-UAV enabled secure communication system via jointly optimizing the UAV trajectories and user scheduling. Reference \cite{Hua} optimizes the UAV trajectory, transmit power and user scheduling to maximize the achievable secrecy rate per energy consumption unit in UAV-enabled secure communications. References \cite{Sun, Guo, Jiang} study more complicated UAV enabled communications systems with 3-D trajectory.

Due to the limited endurance and on-board energy of UAVs, the problem of UAV energy minimization has attracted much attention \cite{Z_Cheng,Y_Zeng_1,Sambo,Zhan,Song}. The work \cite{Sambo} applies the genetic algorithm to design the trajectory with the least energy consumption to visit all BSs and return to the UAV station. Reference \cite{Song} minimizes the completion time and energy consumption problems for a fixed-wing UAV-enabled multicasting system via jointly optimizing the flying speed, UAV altitude, and antenna beamwidth. In \cite{Z_Cheng}, the authors consider the joint problem of the sensor nodes' wake-up schedule and the trajectory to minimize the maximum energy consumption while guaranteeing the reliability of the data collected from the sensors. Nevertheless, these works did not consider UAV's propulsion energy consumption, which is important for UAV's lifetime. In \cite{Y_Zeng_1}, the authors derive a closed-form propulsion power consumption model for rotary-wing UAVs. Then, by using this model, they aim at minimizing the total energy consumption via joint optimization of trajectory and time scheduling between GUs. Based on the energy model in \cite{Y_Zeng_1}, \cite{Zhan} minimizes the maximum energy consumption of all Internet-of-Things (IoT) devices while complying with the energy budget requirement.

Recently, there has been a growing research interest in applying dynamic programming (DP) in UAV-enabled wireless communications \cite{JGong,Hu}. The authors in \cite{JGong} solve the problem of flight time minimization for data collection in a one dimensional wireless sensor networks (WSNs). More specifically, the DP algorithm is proposed to find the optimal data collection intervals of multi-sensors. In \cite{Hu}, the problem of optimizing the spectrum trading between macro base station (MBS) manager and UAV operators is solved. Then, the DP is adopted to find the optimal bandwidth allocation which is the most suitable for each UAV operator. To the best of our knowledge, there are no other works consider DP to solve the problem of \textit{UAV trajectory design with latency constraints.}

The aforementioned works have addressed the various new challenges in UAV-enabled communications, such as completion time minimization \cite{Y_Zeng}, energy minimization \cite{Z_Cheng,Y_Zeng_1,Sambo,Zhan,Song}, and throughput maximization \cite{Jiang}. Moreover, efficient methods have been devised to deal with complicated optimization problems, e.g., time discretization method \cite{Z_Cheng,Guo}, path discretization method \cite{Y_Zeng_1}, block coordinate descent (BCD) in combination with the successive convex approximation (SCA) method \cite{Zhan}, and efficient trajectory design \cite{Y_Zeng}. Specifically, \cite{Y_Zeng,Y_Zeng_1} and \cite{Zhan} have proposed a new framework to design an efficient trajectory by applying TSP solution. Basically, the TSP asks the question of finding the shortest path that visits all users in the network and returns to the origin point which is an NP-hard problem in combinatorial optimization. Thus, a joint problem of trajectory design and other communications factors (e.g., communication scheduling,  transmit power allocation, time allocation) in \cite{Y_Zeng_1,Zhan} is even more challenging. In order to overcome these problems, the authors of \cite{Y_Zeng_1,Zhan} are wisely using TSP solution as an initialized feasible trajectory in their proposed iterative algorithms. Despite remarkable achievements, none of works in \cite{Y_Zeng,Y_Zeng_1,Zhan} take the time constraints into consideration. 

To overcome the limitation in \cite{Y_Zeng,Y_Zeng_1,Zhan}, our work studies the UAV-enabled communications systems in practical scenarios in which the GUs' transmissions are subject to some latency or requested timeout (RT) constraints. The considered system is motivated from realistic communication-related applications, e.g., content delivery networks \cite{Vu2} or the age of information or data collection, in which when a GU requests content data, it needs to be served within a certain RT. For example, in an emergency case or during a natural disaster, data need to be collected/transmitted promptly for evaluations/disseminations of the current situation in a given area. Concretely, the data from sensor nodes with limited storage capacity need to be collected in time for the continuous measurements before it becomes useless or being overwritten by incoming data. Besides that, the vital information must be disseminated to people about incoming disaster as soon as possible. Depending on the important role of each region, the different requested timeout values will be assigned for each area. Our goal is to design an energy-efficient UAV trajectory while guaranteeing the predefined RT constraints of all GUs. It can be seen that the considered system is clearly different from \cite{Y_Zeng,Y_Zeng_1,Zhan}. Therefore, the TSP-based method in those works can not be directly applied in this paper. This motivates us to propose a new approach to solve problem of UAV trajectory design for energy minimization with latency constraints. Concretely, we propose trajectory design algorithms based on TSPTW which is a generalization of the classic TSP and has applications in many important sequencing and distribution problems \cite{Dumas}. The TSPTW requires that each node (or user) must be visited within a predefined time window. The time window includes start time and end time (or requested timeout) associated with each node in the network. The start time and end time define when the service at the considered node can begin and finish. In summary, our contributions are as follows: 

\begin{itemize}
        \item Firstly, we find a feasible set of paths while satisfying the RT constraints for all GUs. In order to deal with the nature NP-hardness of the formulated problem, we propose two algorithms, namely, \emph{DP}, and \emph{heuristic} algorithms based on the TSPTW and they are compared with {\emph{exhaustive search}} and TSP-based method \cite{Y_Zeng,Y_Zeng_1}. While the exhaustive search algorithm provides the global optimality, its exponential computation complexity might limit its applicability in practical applications. In such cases, the heuristic algorithm with a lower complexity is often considered to be a suitable replacement. However, this solution significantly decreases the performance compared to that of exhaustive search. Thus, DP is proposed as a new algorithm to balance between exhaustive and heuristic algorithms. Especially, its performance converges to that of exhaustive at a much lower complexity.
          
        \item Secondly, we minimize the total UAV's energy consumption for each given path in a feasible set via a joint optimization of the UAV velocities in all hops. Since the formulated problem is proved to be convex, it can be solved by using standard methods. Then, the path with lowest energy consumption which also satisfies the energy budget constraint is selected as a designed trajectory for UAV. Notably, in this work all the computation for path design is performed in an offline manner, i.e., prior to the UAV flight. 
        
        \item Finally, the effectiveness of the proposed algorithms is demonstrated via numerical results, which show significant improvements in both energy consumption and outage probability compared with our benchmarks \cite{Y_Zeng,Y_Zeng_1}.
\end{itemize}

The rest of this paper is organized as follows. Section \ref{System_model} introduces the system model. The energy-efficient UAV communication with path design and velocity optimization is analyzed in Section \ref{Sec:E}. Section \ref{Sec:Num} shows the simulation results. Finally, discussion and concluding remarks are given in Section \ref{Sec:Con}.

\textit{Notations:} Scalars and vectors are denoted by lower-case letters and boldface lower-case letters, respectively. For a set $\mathcal{{K}}$, $|\mathcal{{K}}|$ denotes its cardinality. For a vector $\bf v$, $\left\|\bf v \right\|$ denotes its Euclidean ($\ell_2$) norm. $C_k^K$ denotes a set of all $k$-combinations of $K$ elements in set $\cal K$. The notion $\bx\preccurlyeq \by $ means each element of vector $\bx$ is smaller than $\by$. $\mathbb{E}\{x\}$ denotes the expected value of $x$. $\mathbb{P}\{a\}$ is representing the probability for happening event $a$.%$\| \cdot\|$ represents the Euclidean ($\ell_2$) norm  of a vector.  $1_{m,n}$ is the $m \times n$ matrix of ones where every element equals to one. $\mathbb{R}^+$ represents for the nonnegative real numbers, i.e., $\mathbb{R}^+=\{x \in \mathbb{R}|x \ge 0\}$. $\mathbb{R}^{++}$ represents for the positive real numbers, i.e., $\mathbb{R}^{++}=\{x \in \mathbb{R}|x > 0\}$.
%%%%%%%%%%%%%%%%%%%%%%%%%%%%%%%
% FIG1 FIG1 FIG1 FIG1 FIG1 FIG1  
\begin{figure}[t]
       \centering
        \includegraphics[width=9cm,height=7cm]{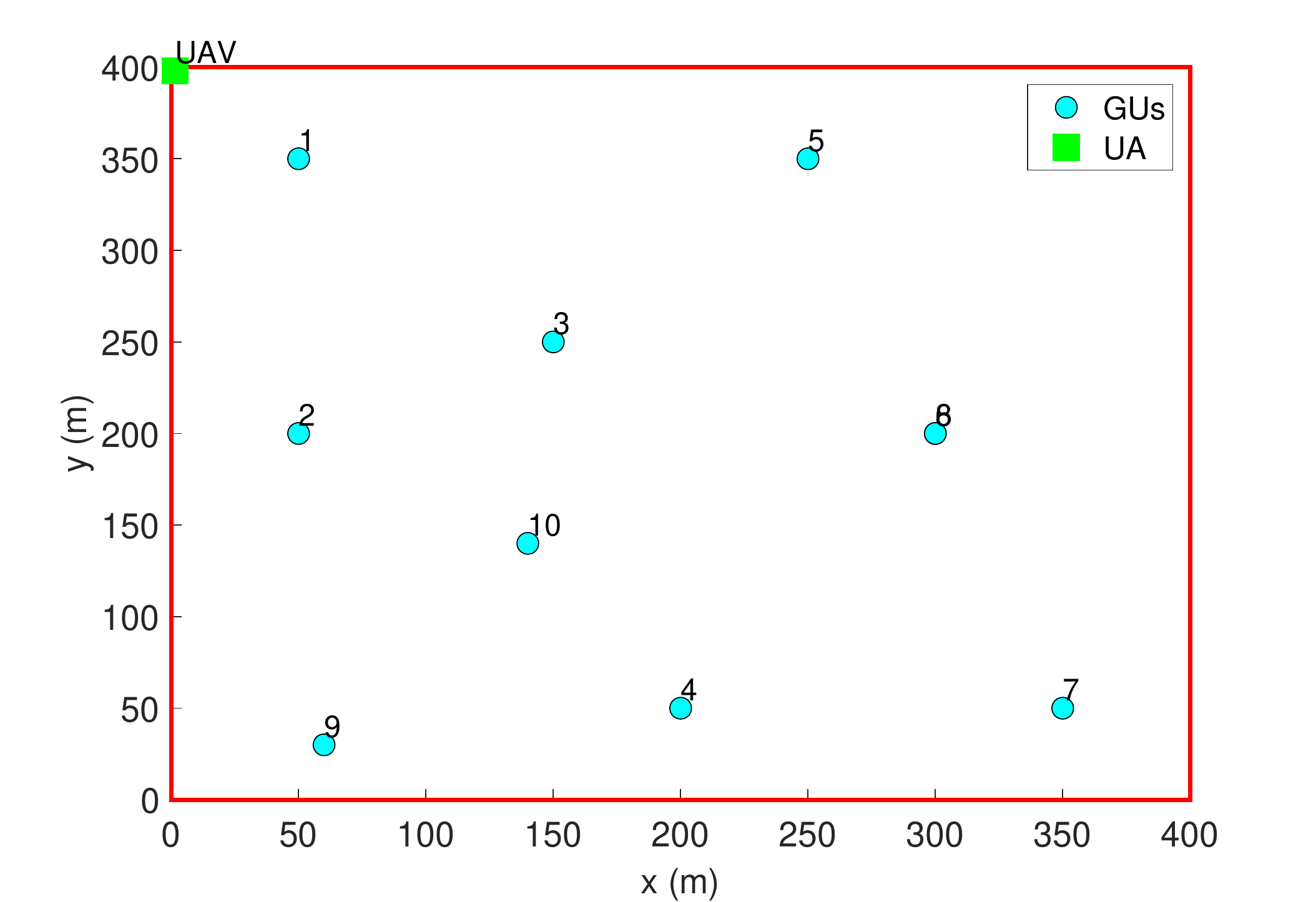}
        \caption{System model.}
        \label{fig:1}   
\end{figure}
% FIG1 FIG1 FIG1 FIG1 FIG1 FIG1
%%%%%%%%%%%%%%%%%%%%%%%%%%%%%%%%%%%%%%%%%%%%%%%%%%%%%%%%%%%%%%%%%%%%%%
%                              SYSTEM MODEL                          % 
%%%%%%%%%%%%%%%%%%%%%%%%%%%%%%%%%%%%%%%%%%%%%%%%%%%%%%%%%%%%%%%%%%%%%%

%\newpage
\section{System Model} 
\label{System_model} We consider an UAV-enabled communication system in which a UAV helps to transmit data to a set of $K$ ground users (GUs), denoted by $\mathcal{K}\triangleq \{1,\ldots,K\}$. Due to limited access, the users can only receive data from the UAV \cite{Y_Zeng,Y_Zeng_1}. The location of GU $k$ is denoted as ${\bq_k} \in {\mathbb{R}^{2 \times 1}}, k \in \mathcal{K}$. Let $(u_1, u_2, \dots, u_K)$ be a permutation of $(1, 2, \dots, K)$, and let $\bu \triangleq [u_1\; u_2\; \cdots\; u_K]^T$ specify a trajectory of the UAV to serve all users following the path $ {0 \to {u_1} \to {u_2} \to ... \to {u_K} \to 0} $, where $0$ denotes the UAV station (or depot).  It is assumed that GU\ $k$ is required to be served within $\eta_k$ units of time after the start of the UAV's mission. We refer to $\eta_k$ as the requested timeout of GU\ $k$, for $k \in {\cal K}$. 

\subsection{Trajectory Design Model}
In literature, there are different trajectory design models for UAV communications, e.g., Table \ref{review:table}. Basically, it can be classified into two types such as coarse (e.g., hovering-communications \cite{Y_Zeng}) and fine trajectory design (e.g., fly-hover-communication (FHC) and flying-and-communication (FAC) methods in \cite{Y_Zeng_1}, virtual base station (VBS) as waypoints (VAW) and waypoints based on VBS placement and convex optimization (WVC) methods in \cite{Y_Zeng}). In hovering-communications method, the UAV has to move to GU $k$'s location and keeps hovering during the transmission period. The authors in \cite{Y_Zeng, Y_Zeng_1} utilize TSP to find the visiting order of $K$ GUs' locations. Based on this result, the FHC method optimizes the visited locations, each  for communicating with one GU, instead of hovering over each GU. The VAW is similar to FHC method, the difference is that each hovering location in VAW is for communicating with a group of GUs. In \cite{Y_Zeng}, FAC method, an updated version of FHC, is proposed in which the UAV can communicate while flying. In \cite{Y_Zeng_1}, the authors improve VAW method by proposing a more efficient waypoint (i.e., hovering point) design to reduce the traveling distance, i.e., WVC method. Based on these examples, we can conclude that the fine trajectory design can be obtained based on the coarse one.

In this work, hovering-communication is applied as the trajectory design model since it is a very intuitive protocol that is also easy to implement in practice. Fig. \ref{fig:1} depicts a two-dimensional Cartesian coordinate system, whereas the UAV is located at the ground station and the GUs are located in the considered area.

%%%%%%%%%%%%%%%%%%%%%%%%%%%%%%%%%%%%%%%%%%%%%%%%%%%%%%%
\begin{table*}[h]
	\caption{Trajectory design methods}
	\label{review:table}
	\centering
	\small
	\begin{tabular}{|p{4cm}|p{4cm}|p{8cm}|}
		\hline
		\textbf{Name} & \textbf{No. of visited GUs per 1 hovering point}& \textbf{Hovering point design}\\
		\hline
		Hovering-communications \cite{Y_Zeng}& 1 & GU's location as hovering points\\
		\hline
		FHC \cite{Y_Zeng_1} & 1 & Hovering point is within the GUs' transmission range\\
		\hline
		FAC \cite{Y_Zeng_1} & 1 & Hovering point is within the GUs' transmission range and UAV transmits data when flying\\
		\hline
		VAW \cite{Y_Zeng} & Multi-GUs & Hovering point is the VBS defined as the center of a group of GUs\\
		\hline
		WVC \cite{Y_Zeng} & Multi-GUs & Based on VBS in VAW, they optimize the hovering point using optimization method\\
		\hline
	\end{tabular}
\end{table*}
 
\subsection{Transmission Model}
The UAV's trajectory is split into $K+1$ line segments (or hops) which are represented by all connections between $K+2$ way-points on any given route {(see Fig.~\ref{fig:2} for details).} We assume that the UAV flies at a constant altitude of $H$ (meters). Therefore, the distance traveled from GU\ $j$ to GU\ $k$ is given by %
\begin{align}
\label{eq:distance}
l_{j\to k} =  \| \bq_j - \bq_k\|, 0\le j,k \le K+1,
\end{align}
where the {index} $0$ represents the UAV station. We assume that the UAV velocity is constant during each hop but can change from hop to hop.

For $i =1,2, \ldots K+1$, let $v_i$ denote the UAV velocity at the $i$-th hop, while for $ k = 1, 2,\ldots, K$,  $\tau_k$ stands for  the transmission time needed for UAV to send the requested data stream to GU $k$ reliably. Then, for a given trajectory signified by $\bu$, the time for the UAV to reach the  GU $u_k$ is calculated as
\begin{align}
T_k = \sum_{i=1}^{k} \left(t_{u_i}+ \tau_{u_i} \right),\;\; \mbox{ for }  1 \le k \le K,
\end{align}
where $t_{u_i} \triangleq  \displaystyle  \frac{d_i}{v_i}$  and $d_i = l _{u_{i-1} \to u_{i}} $ represent the travel time  and the distance in the $i$-th hop, respectively, for $i = 1,2, \ldots ,K+1$. 

We assume the channel between the GU and the UAV follows a Rician fading \cite{Azari,Gong}, where channel coefficient between GU $k$ and UAV, $h_{k}$, can be written as 
\begin{align}
\label{eq:2_1}
h_{k} = \sqrt {\mu_k }  g_{k},
\end{align}
where $\mu_k $ represents for the large-scale average channel power gain accounting for signal attenuation including pathloss and shadowing and $g_{k}$ accounts for small-scale fading coefficient.

In particular, $\mu_k $ can be modeled as
\begin{align}
\label{eq:2_11}
\mu_k = \mu_0 H^{-\alpha},
\end{align}
where $\mu_0$ is the average channel power gain at the reference distance, and $\alpha$ is the path loss exponent. Then, the small scale fading $g_k$ with expected value $\mathbb{E} \left[| g_k|^2 \right]=1$, is given by
\begin{align}
\label{eq:2_12}
g_k= \sqrt{\frac{G}{1+G}} \overline{g}_k + \sqrt{\frac{1}{1+G}} {\mathop g\limits^\sim}_k,
\end{align}
where $G$ is the Rician factor; $\overline{g}_k$ denotes the deterministic LoS channel component; ${\mathop g\limits^\sim}_k \sim \mathcal{CN}(0,1)$ denotes the Rayleigh fading channel accounting for NLoS components. Then, the maximum achievable
rate between the UAV and GU $k$ is calculated as
\begin{align}
\label{eq:3_0}
R_k  =  B \log_2 \left(1 + \Upsilon |g_k|^2  \right), 
\end{align}
where $\Upsilon \triangleq  \frac{P_{com} \mu_0}{H^\alpha \sigma^2} $, $B$ is the channel bandwidth, $P_{com}$ is the transmit power of the UAV, and $\sigma^2$ is the noise power. 

\begin{figure}[t]
	\centering
	\includegraphics[width=9cm,height=7cm]{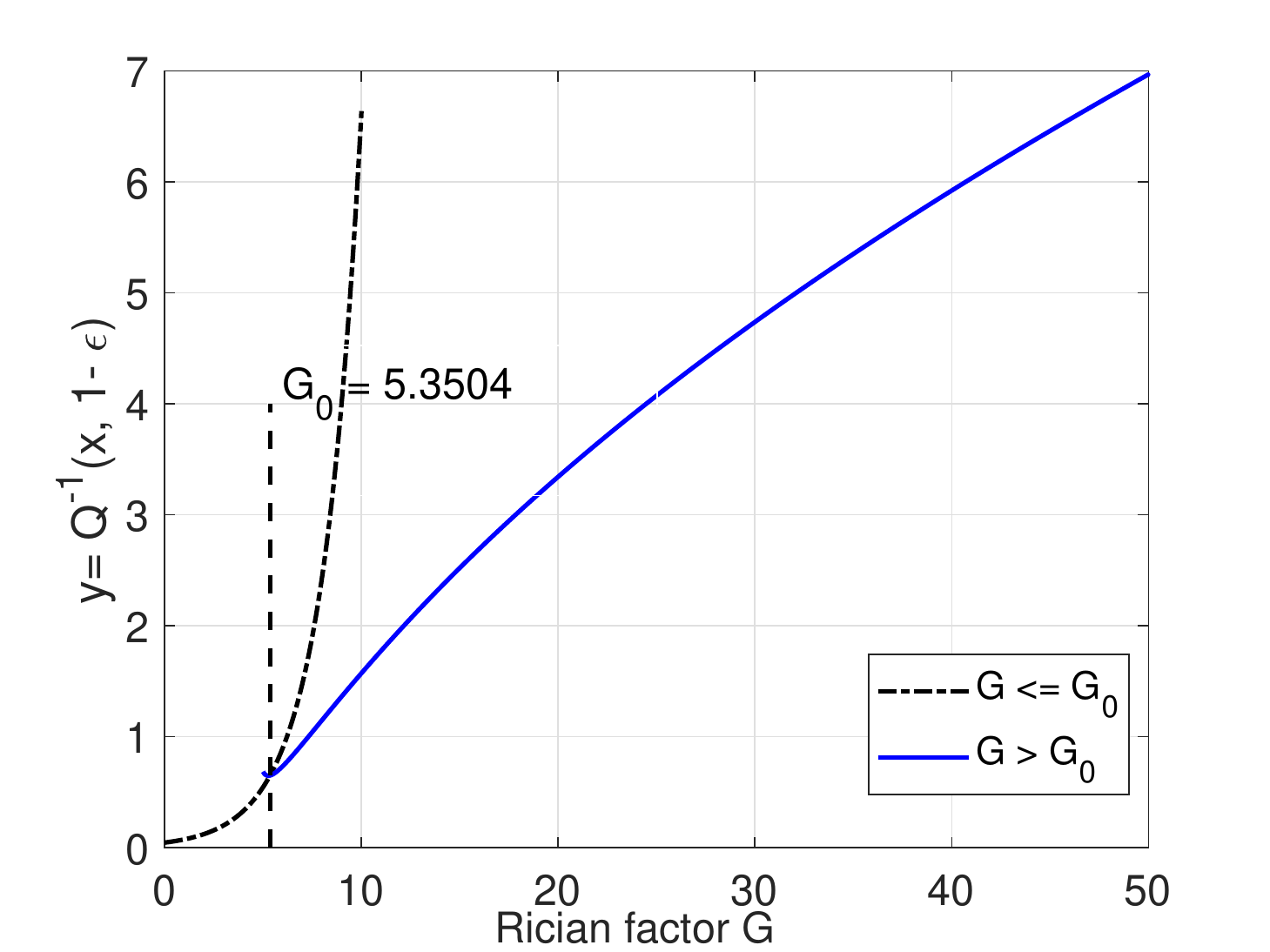}
	\caption{ $y_Q$ curves corresponding to its sub-functions, with $\epsilon=0.001$ .}
	\label{fig:11}   
\end{figure}

As the lack of the knowledge for instantaneous channel state information (CSI) prior to the UAV's flight, the rate $R_k$ is not exactly known. Therefore, the approximated rate for GU $k$ is adopted, i.e., $\overline{R}_k$. Specially, $\overline{R}_k$ is chosen so that $\mathbb{P}\{R_k < \overline{R}_k\}$ remains below or equals to a target $\epsilon$. Moreover, the outage probability that the
GU $k$ cannot successfully receive the transmitted data from UAV, i.e., $\mathbb{P}\{R_k < \overline{R}_k\}$, is expressed mathematically as follows \cite{Azari}:
\begin{align}
\label{eq:31}
& \mathbb{P}\{R_k < \overline{R}_k\} \notag \\ & = \mathbb{P} \Big\{ |g_k|^2 < \frac{ \big( 2^{\overline{R}_k/B}-1 \big)}{\Upsilon} \Big\}  \notag \\
 & = 1 - Q_1 \Big\{x_Q, y_Q \Big\} \le \epsilon,
\end{align}
where $x_Q \triangleq \sqrt{2K}$, $y_Q \triangleq  \sqrt{2 \big( 2^{\overline{R}_k/B}-1 \big) (1+G)/\Upsilon} $, $Q_1(x,y)$ is the first order Marcum Q-function. Moreover, at maximum tolerable value of $\epsilon$, i.e., $Q_1 \Big\{x_Q, y_Q \Big\} = 1 - \epsilon$, $y_Q$ is defined as \cite{Azari}  
\begin{subnumcases} {\label{eq:32} y_Q =}
\sqrt{-2 \log (1-\epsilon)} e^{G/2}, \hfill G \le G_0 \label{eq:32a}\\
\sqrt{2G} + \frac{1}{2 Q^{-1}(\epsilon)} 
\log \big(\frac{\sqrt{2G}}{\sqrt{2G}-Q^{-1}(\epsilon)} \big)  \label{eq:32b}\\
\sqrt{2G} + \frac{1}{2 \sqrt{2G}}, \hfill G > G_0\; \& \; Q^{-1} (\epsilon) = 0 \label{eq:32c}
\end{subnumcases} 
where $G_0$ is the intersection of sub-functions at $\sqrt{2G} < max [0,Q^{-1}(\epsilon)]$ and $Q^{-1}(x)$ is the inverse Q-function. Fig. \ref{fig:11} shows that $G_0$ can be obtained graphically based on the $y_Q$ sub-functions. 

%% FIG2 FIG2 FIG2 FIG2 FIG2 FIG2  
%% FIG2 FIG2 FIG2 FIG2 FIG2 FIG2  

In this work, since UAV hovers right above the user, we can consider a Rician channel with strong LoS component (i.e., high $G$ factor) and $Q^{-1}(\epsilon) \ne 0$. Thus, the $y_Q$ function can be approximated as $y_Q=\sqrt{2G} + \frac{1}{2 Q^{-1}(\epsilon)} 
\log \big( \frac{\sqrt{2G}}{\sqrt{2G}-Q^{-1}(\epsilon)} \big) - Q^{-1}(\epsilon)$. This yields the approximated rate $\overline{R}_k$ can be expressed as:
\begin{align}
{\overline{R}_k} = B \log_2 \left(1+ \frac{{y_Q}^2 \Upsilon}{2(1+G)} \right), 
\end{align}

We adopt the Rician model because it can capture both LoS and NLoS links.

%%%%%%%%%%%%%%%%%%%%%%%%%%%%%%%%%%%%%%%%%%%%%%%%%%%
\subsection{Energy Consumption Model}\label{sec:energy}
The energy consumption of the UAV consists of two types: propulsion energy consumption and communication energy consumption. The former measures the energy consumed to fly or hover over the UAV. The latter is used to transmit data to the GUs. In general, the energy consumption depends not only on the UAV velocity, but also on its acceleration/deceleration. Note that the energy consumption during UAV's acceleration/deceleration is ignored in \cite{Y_Zeng_1,Zhan}  which is reasonable for scenarios when the acceleration/deceleration speed or acceleration/deceleration duration
is small. The power consumption of a rotary-wing UAV flying at velocity $v$ is given as \cite[Eq.~(12)]{Y_Zeng_1}
% EQ5
\begin{align}
\label{eq:5}
P_{\rm fly}(v) =& \underbrace {{P_0}\left( {1 + {\alpha _1}{v^2}} \right)}_{{\rm{blade profile}}} \notag \\
&+ \underbrace {{P_1}\sqrt {\sqrt {1 + \alpha_2^2 v^4}  - {\alpha _2}{v^2}}}_{{\rm{induced}}} + \underbrace {{\alpha _3}{v^3}}_{{\rm{parasite}}},
\end{align}
% EQ5
where ${P_0} = \frac{\delta }{8}\rho sA{\Omega ^3}{R^3}$, ${P_1} = (1 + {I})\frac{{{W^{3/2}}}}{{\sqrt {2\rho A} }}$, ${\alpha _1} = \frac{3}{\Omega ^2 R^2},$ ${\alpha _2} = \frac{1}{2V_R^2},$ and ${\alpha _3} = 0.5{a_0}\rho sA$. Blade profile power, parasite power, and induced power are needed to overcome the profile drag of the blades, the fuselage drag, the induced drag of the blades, respectively. Other parameters are explained as in Table I of \cite{Y_Zeng_1}. 

\begin{figure}[t]
	\centering
	\includegraphics[width=9cm,height=7cm]{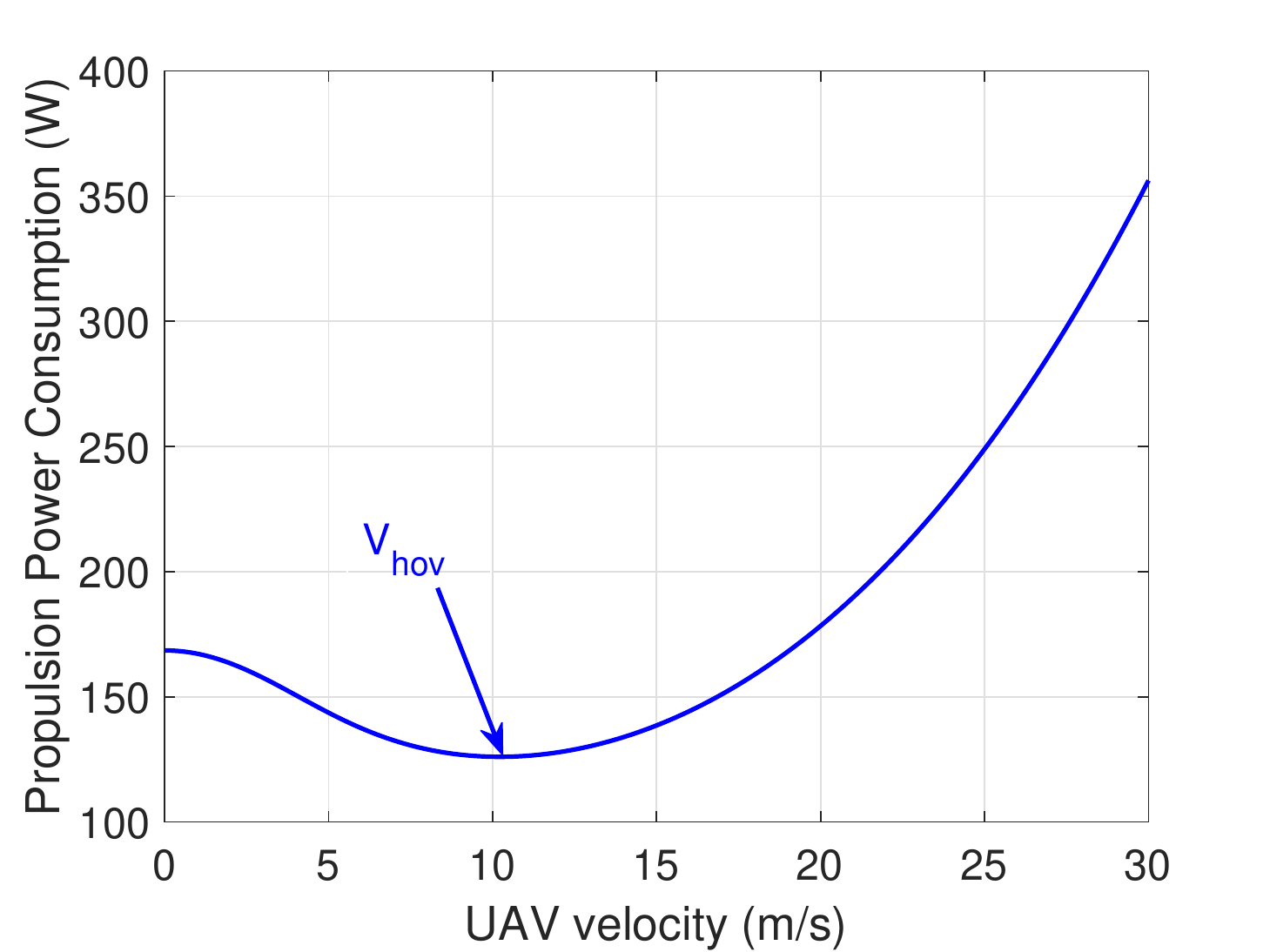}
	\caption{Propulsion power consumption versus velocity.}
	\label{fig:13}   
\end{figure}

For $i= 1,2\ldots ,K$+1, the total energy consumption that the UAV spends on hop $i$ is given as
\begin{align}
\label{eq:6}
E_i(v_i, d_{i}) = E_{{\rm fly},i} (v_i, d_i)+ E_{{\rm hov},i} + E_{{\rm com},i},
\end{align}
where $E_{{\rm fly},i} (v_i, d_i)= P_{\rm fly}(v_i)\times t_{u_i} = P_{\rm fly}(v_i)\times d_i/v_i$, $E_{{\rm hov},i} = P_{\rm fly}(v_{\rm hov})\times \tau_{u_i}$, and $E_{{\rm com},i} = P_{\rm com}\times  \tau_{u_i}$ are the energy consumption due to flying, hovering, and communications, respectively, where $P_{\rm fly}(v_i)$ is provided in \eqref{eq:5}. When UAV approaches the GU, it will fly around the GU with certain velocity $v_{\rm hov}$ instead of hovering directly above it to minimize the energy consumption \cite{Y_Zeng_1}. Moreover, the energy consumption due to hovering is $E_{{\rm hov},k} = P_{\rm fly}(v_{\rm hov})\times \tau_{k}$, where $P_{\rm fly}(v_{\rm hov})$ and $\tau_k$ are the propulsion energy consumption due to hovering and transmission time to serve GU $k$, respectively. Furthermore, $\tau_k$ is computed as $\tau_k = Q_k/\overline{R}_k$, where $Q_k$ denotes the length of the requested content in bits and $\overline{R}_k$ denotes the approximated transmission rate from the UAV to GU $k$. Since $Q_k$ and $\overline{R}_k$ can be obtained prior to the UAV flight. Thus, $E_{{\rm hov},k}$ is proportional to $P_{\rm fly}(v_{\rm hov})$ which is minimized at $v_{hov}$ value as in Fig. \ref{fig:13}.
%% FIG3 FIG3 FIG3 FIG3 FIG3 FIG3   
%% FIG3 FIG3 FIG3 FIG3 FIG3 FIG3 

%%%%%%%%%%%%%%%%%%%%%%%%%%%%%%%%%%%%%%%%%%%%%%%%%%%%%%%
\section{Energy-efficient UAV communication with path and velocity optimization}
\label{Sec:E}
Our goal is to jointly design the path and velocities to minimize the total energy consumption while satisfying the RT constraints and energy budget for all GUs. Intuitively, we aim to find the visiting order $\mathbf{u} \triangleq [u_1, \dots, u_K]$ and the UAV velocities which result in the smallest energy consumption. Then, the problem is formulated as
\begin{align}
\label{OP:energy}
\mathcal{P}_1: &\min_{\mathbf{u},\{v_i\}_{i=1}^{K+1}} ~~ \sum\limits_{i = 1}^{K+1} ({E_{{\rm fly},i} (v_{i} , d_{i})+ E_{{\rm hov},i} + E_{{\rm com},i}})  \\
\mathtt{s.t.}~~
&\text{C1}:~\sum_{i=1}^{k} \left(\frac{d_i}{v_i} + \tau_{u_i} \right) \le \eta_{u_k},\;\; \mbox{ for } \;\; 1\le k \le K \notag\\
&\text{C2}:~0 \le v_i \le V_{\rm max},  \;\;\; \mbox{ for }  1\le i \le K+1, \notag\\
&\text{C3}:\sum\limits_{i = 1}^{K+1} ({E_{{\rm fly},i}(v_i,d_i) + E_{{\rm hov},i} + E_{{\rm com},i}})\le E_{\rm tot}, \notag \\
&\text{C4}: |v_{i+1} - v_{i}| \le \Delta V, \forall 1 \le i \le K. \notag
\end{align}

In $\mathcal{P}_1$, constraint C1 guarantees the RT requirement for the GUs which states that the maximum latency to serve GU $u_k$ cannot exceed the predefined RT $\eta_{u_k}$, C2 requires that the flying speed of the UAV must be less than the maximum velocity $V_{\rm max}$, and C3 means that the total energy consumption of UAV on the considered path should not exceed the total energy budget $E_{\rm tot}$. Otherwise, this is an infeasible path. C4 guarantees that the traveling speed of UAV between two consecutive hops is less than a predetermined value. 

Problem \eqref{OP:energy} requires optimizing the path $\bu$ and traveling velocities $\{v_i\}_{i=1}^{K+1}$ of the UAV on all hops. Note that \eqref{OP:energy} includes a complicated energy consumption as in C3, as well as an objective function which depends on the designed path $\bu$. However, as the objective function is the same of the LHS in C3, we can, without loss of generality, solve ${\cal P}_1$ without C3 and find the minimum total energy consumption. If this energy is less than $E_{\rm tot}$, then ${\cal P}_1$  is feasible and its solution is the same as the solution to ${\cal P}_1$ without C3, otherwise ${\cal P}_1$ is not feasible and we say {\em outage} has occurred. 

\begin{definition}
	The OP is defined as $\mathbb{P}\{R_k < \overline{R}_k\}$ and the probability that no feasible path (a path that satisfies all the GUs' RT requirements and $E_{tot}$) is found.
\end{definition}

Note that even without C3, problem ${\cal P}_1$ is a TSPTW, itself is already NP-hard \cite{Dumas}. To solve this problem, we first find the feasible set of paths (denoted as $\mathcal{U}^\star$) which satisfy constraints C1 while  the hop velocities satisfy C2. Next, we will minimize the energy consumption on each given path via joint optimization of velocities over all hops. Finally, the lowest energy consumption path which satisfies the energy budget constraint C3 is chosen as the trajectory design for UAV.
        
\subsection{Obtain a feasible set of paths}
In this section, we introduce three solutions, namely, exhaustive search, heuristic search, and DP algorithms to obtain a feasible set of paths that satisfy constraints C1, C2 and C4. The exhaustive search gives the best solution with very high complexity. The heuristic tries to reduce the complexity but the performance also decreases. Thus, the DP is proposed as a solution to balance between exhaustive search and DP algorithms.

Specifically, the  feasible set  of all paths which satisfy constraints C1, C2, and C4, i.e., $\mathcal{U}^\star$, will be obtained by choosing $v_i=V_{\max}$, for $i=1,2,\ldots,K+1$.  

The proof of this property relies on the monotonically decreasing behavior of the LHS of constraint C1 with respect to $v_i$, for any given $i$. For any $k \in {\cal K}$, if a path satisfies C1 with $v_i < V_{\rm max}$, then this path satisfies C1 with $v_i = V_{\rm max}$, as the LHS of C1 is  monotonically decreasing in $v_i$, for $i=1,2,\ldots,K$. Also, for any $k \in {\cal K}$, if a path does not satisfy C1 with $v_i = V_{\rm max}$, then this path does not satisfy C1 with $v_i < V_{\rm max}$, as $\displaystyle \sum_{i=1}^{k} \left(\frac{d_i}{v_i} + \tau_{u_i} \right) > \displaystyle \sum_{i=1}^{k} \left(\frac{d_i}{V_{\max}} + \tau_{u_i} \right) > \eta_{u_k}$.  

By considering $v_i=V_{\rm max}$ as above discussion, the following subsections present three proposed algorithms to find a feasible set of paths. 
\subsubsection{Algorithm 1}\textit{Exhaustive search algorithm }

%In this subsection, we introduce an {exhaustive} search algorithm used as the reference in the section Simulation Results. 
The principle of the exhaustive search algorithm is to visit all the paths and find Hamiltonian cycle paths \cite{Bollobas} satisfying the RT constraint. For each path in the feasible set, we minimize the energy consumption via jointly optimizing the velocities as in Section \ref{Subsec:B}. Thus, in order to reduce the computational complexity for solving \eqref{OP:energy}, we only take $\Psi$ feasible paths into consideration.

This problem is in a form of TSPTW problem, which can be solved by finding the minimum cost tour (Hamiltonian cycle path) starting and ending at location $0$ and visiting all GUs only once \cite{Dumas}. The details is summarized in Algorithm 1. Firstly, we initialize all the parameters as in lines 1 and 2. More specifically, $\bq_k$, $\tau_k$, $\eta_k$ are the location, data transmission time, requested timeout constraint of GU $k$, respectively; $\cal U$ is the set containing all feasible paths; $l_{0 \to k}$ and $a_{0k}$ are the traveling distance from UAV to GU $k$ and the total time needed for UAV to finish transmitting data to GU $k$, respectively.
%%%%%%%%%%%%%%%%%%%%%%%%%%%%%%%%%%%%%%%%%%%%%%%%
\begin{algorithm}[t]
        \caption{Exhaustive search algorithm for solving $\mathcal{P}_2$} 
        \label{alg:1}
        \begin{algorithmic}[1]
                \State {\textbf{Input:} $V_{\rm max},$ $\{\bq_k, \tau_k, \eta_k\}_{k=1}^K$.}
                \State {\textbf{Initialize}: Calculate the set $\mathcal{I}$ containing all the paths, $\boldsymbol{\eta_0} = [\eta_1, \dots, \eta_K]$, $\mathcal{U}=\O$, $l_{0 \to k}=\|{\bf q}_k - {\bf q}_0\|, a_{0k}=\frac{l_{0 \to k}}{V_{\rm max}} + \tau_{k}, k\in {\cal K}$,  ${\bf a}=[a_{01},\dots,a_{0K}]$.}
                \If {${\bf a} \le \boldsymbol{\eta_0}$}  \Comment {Check feasibility}
                \For {$m = 1: |\mathcal{I}|$} \Comment{For each path $\mathbf{u}^{(m)} \in \mathcal{I}$}
                \State {$L^{(m)}_k = \sum\limits_{i=1}^{k} \left(\frac{l_{u^{(m)}_{i-1} \to u^{(m)}_{i}}}{V_{\rm max}} + \tau_{i} \right)$ \Comment {Total traveling time between GUs $u_{i-1}$ and $u_i$ in $\mathbf{u}^{(m)}$ and data transmission time to $u_i$}}
                \If {$\bL^{(m)}\preccurlyeq\ \boldsymbol{\eta_0}$} \Comment {Check the RT constraint}
				\State $\mathcal{U}=\mathcal{U} \cup \mathbf{u}^{(m)}$ \Comment Update feasible paths
                \EndIf  \textbf{end if}
                \EndFor \textbf{end for}
                \State Find $\Psi$ shortest feasible paths, i.e., $\mathcal{U}^\star \in \mathcal{U}$
                \EndIf \textbf{end if}
                \State \textbf{Output:} $\mathcal{U}^\star$.
        \end{algorithmic}
\end{algorithm}
%%%%%%%%%%%%%%%%%%%%%%%%%%%%%%%%%%%%%%%%%%%%%%
%
Basically, Algorithm 1 consists of two steps. In the first step, we check the RT constraint from UAV station to each GU $k$, as in line 3. Based on the triangle inequality constraint, if there exists any GU $k$ which does not satisfy the RT constraint, it has no feasible path. Otherwise, we  will try all $K!$ paths which visits all the GUs once, lines 4 to 9. For each path $\mathbf{u}^{(m)} \in {\cal I}$, we calculate the UAV traveling time from $u_{i-1}$ to $u_i$ and data transmission time for $u_i$, with $u_{i-1}$ and $u_i \in \mathbf{u}^{(m)}$. It then compares the visit time to every GU with the corresponding RT requirements (constraint C1 in \eqref{OP:energy}), as in line 6. If path $\mathbf{u}^{(m)}$ satisfies the RT constraint, it will be accumulated to set $\cal U$, line 7. Thus, the complexity of Algorithm 1 is $\mathcal{O}(K!)$ \cite{Lawler}. Finally, the set of $\Psi$ feasible paths which satisfies all the RT constraints and imposes the $\Psi$ shortest traveling time will be selected.

%%%%%%%%%%%%%%%%%%%%%%%%%%%%%%%%%%%%%%%%%%%%%%%%%%%
\begin{algorithm}[t]
	\caption{Heuristic algorithm for solving $\mathcal{P}_2$}\label{alg:2}
	\begin{algorithmic}[1] 
		\State \textbf{Input:} $V_{\rm max},$ $\{\bq_k, \tau_k, \eta_k\}_{k=1}^K$.
		\State \textbf{Initialize}: $\mathbf{U} = \O$, $\mathcal{I}_{+} =\O,$ $\mathcal{I}_{-} ={\cal K},$ $l_{0 \to k}=\|{\bf q}_k - {\bf q}_0\|, a_{0k}=\frac{l_{0 \to k}}{V_{\rm max}} + \tau_{k}, k\in {\cal K}$, ${\bf a}=[a_{01},\dots,a_{0K}]$, $\boldsymbol{\eta_0} = [\eta_1, \dots, \eta_K]$. 
		\If {{${\bf a} \le \boldsymbol{\eta_0}$}} \Comment Check feasibility in the first hop
		\State Find the closet GU having minimum RT value, i.e., $u^\star \in \mathcal{I}_{-}$.
		\State $\mathbf{U} = \mathbf{U} \cup \{u^\star\}$ \Comment If $u^\star$ satisfy constraint C1.
		\State $\mathcal{I}_{+} = \mathcal{I}_{+} \cup \{u^\star\}$, $\mathcal{I}_- = \mathcal{K}\setminus \{u^\star\}$
		\State Repeat steps 4 to 6 until $\mathcal{I}_-=\O$ or no any GU satisfying the C1 constraint.
		\EndIf \textbf{end if}
		\State \textbf{Output:} $\mathcal{U}^\star = \mathbf{U}$.
	\end{algorithmic}
\end{algorithm}
%%%%%%%%%%%%%%%%%%%%%%%%%%%%%%%%%%%%%%%%%%%%%%
\begin{algorithm}[t]
	\caption{DP-based algorithm for solving $\mathcal{P}_2$}\label{alg:3}
	\begin{algorithmic}[1] 
		\State \textbf{Input:} $V_{\rm max},$ $\{\bq_k, \tau_k, \eta_k\}_{k=1}^K$. 
		\State \textbf{Initialize}: $\Xi_1 \buildrel \Delta \over= \{\xi_1\} $, ${\xi _1}=(\{0\},0),$ $C(\{0\},0)=0,$ ${\cal A}=\{a_{jk}\},$ $\mathcal{U}^\star = \O,$ $l_{0 \to k}=\|{\bf q}_k - {\bf q}_0\|, a_{0k}=\frac{l_{0 \to k}}{V_{\rm max}} + \tau_{k}, k\in {\cal K}$,  ${\bf a}=[a_{01},\dots,a_{0K}]$, $\boldsymbol{\eta_0} = [\eta_1, \dots, \eta_K]$, ${\cal B}_1^0=\O$,
		\If {${\bf a} \le \boldsymbol{\eta_0}$}  \Comment{Check feasibility}    
		\State $\Xi_m = \O$
		\For{$({\cal S},j) \in \Xi_{m-1}$} 
		\State Update $\Xi_m = \Xi_m \cup ((\mathcal{S},j)\cup{k},k)$ \Comment{If $C((\mathcal{S},j)\cup{k},k) \le \eta_k$ and $C((\mathcal{S},j)\cup{k},k)$ is the minimum cost of all states $((\mathcal{S},j)\cup{k},k)$}
		\State Update ${\cal B}_m^k=\{j\}$
		\EndFor \textbf{end for}
		\State $m=m+1$
		\State Repeat steps 4 to 7 until $|{\cal S}|=K+1$ or no any GU satisfying the RT constraint.
		\State For each state $({\cal S},k) \in \Xi_{K+1}$, the visiting order $\mathbf{u}_\star$ is obtained by checking for backward from $\Xi_{K+1}$ to $\Xi_1$ based on ${\cal B}_m^k$.
		\State $\mathcal{U}^\star = \mathcal{U}^\star \cup \mathbf{u}_\star$ 
		\EndIf \textbf{end if}
		\State \textbf{Output:}
 $\mathcal{U}^\star$.
	\end{algorithmic}
\end{algorithm}

\begin{table}[t]
	\caption{Illustration for Travel Time between GUs in Heuristic Algorithm}
	\label{table2}
	\centering
	\small
	\begin{tabular}{|p{1cm}|p{1cm}|p{1cm}|p{1cm}|p{1cm}|}
			\hline
			GU & 0 & 1 & 2 &  3\\
			\hline
			0 &$+\infty$& 1 & 1.2 & 1.3 \\
			\hline
			1 & $1$& $+\infty$& 0.5 & 1.2 \\
			\hline
			2 & 1.2& 0.5 & $+\infty$ & 2 \\
			\hline
			3 & 1.3& 1.2 & 2 & $+\infty$ \\
			\hline
	\end{tabular}
\end{table}
\subsubsection{Algorithm 2}  \textit{Heuristic algorithm}

Although providing near-optimal performance, the high computation complexity of Algorithm 1 may limit its potential in realistic scenarios. In this subsection, we propose a heuristic search algorithm, which compromises the performance against complexity. The key idea behind the heuristic algorithm is to restrict the search space at each step, in which it only foresees one hop ahead when checking the RT condition. Details of the heuristic algorithm are described in Algorithm~\ref{alg:2}. 

Firstly, we initialize all the parameters as in lines 1 and 2. More specifically, $\bf U$ is the feasible path; we can check other parameters as in Algorithm 1. The searching in the heuristic algorithm consists of $K$ steps, in which it maintains two sets: a set of visited GUs and another set of GUs which have not been visited, i.e., $\mathcal{I}_{+}$ and $\mathcal{I}_-$, respectively. First, we check the RT constraint for the first step (or hop) as in Algorithm 1, line 3. If the RT constraint from UAV station to each GU $k$ is satisfied, then, we select the closet GU having minimum RT value as the first visited GU into the designed path, i.e., ${\bf U}$ as in lines 4 to 6. Then, we continue checking a until the set $\mathcal{I}_-$ is empty. In the other hand, if there has no GUs satisfying the RT constraint at the $k$-th step, it is not possible to find out the feasible path. As shown in the Algorithm 2, the fundamental operations employed in the computation are additions and comparisons. The total number of operations needed to run Algorithm 2 from steps 1 to $K$ is $\sum\limits_{k=2}^K \left(K-k+1\right)= \frac{K(K+1)}{2}$ \cite[Eq. (0.121.1)] {Gradshten_2007}. Thus, the complexity of the heuristic algorithm is $\mathcal{O}(K^2)$, which is significantly smaller than the complexity $\mathcal{O}( K!)$ of Algorithm~\ref{alg:1}.

To make it easy to understand, we would like to give an example of heuristic algorithm.  More specifically, we consider ${\cal K}=\{1,2,3\}$, ${\boldsymbol \eta} = \{2,2,5\}$ seconds, $\tau_k=0.12$ seconds, $0$ denotes the UAV station, Table \ref{table2} is the travel time between GUs. Firstly, the vector ${\bf a} = [1.12, 1.32, 1.42]$ can be determined based on the equation $a_{0k}=\frac{l_{0 \to k}}{V_{\rm max}} + \tau_{k}$ as in step 2. Then, the RT constraint can be checked as in step 3. Since all GUs $k$ satisfy the constraint C1, thus, we update the feasible path ${\bf U}= 1$ and $\mathcal{I}_{+}={1}$ and $\mathcal{I}_-=\{2,3\}$ as in steps 4 and 5. Next, the accumulated traveling of UAV to next GUs in $\mathcal{I}_-$ can be computed as $a_{01} + a_{12}=1.12 + 0.62 = 1.74< \eta_2$, $a_{01} + a_{13}=1.12 + 1.32 = 2.44 < \eta_3$. As $a_{01} + a_{12}< a_{01} + a_{13}$, we have ${\bf U}=1 \to 2$, $\mathcal{I}_{+}=\{1,2\}$ and $\mathcal{I}_-=\{3\}$. Next, we check the RT constraint to GU 3, i.e.,  $a_{01} + a_{12} + a_{23}=1.74 + 2.12 = 3.86< \eta_3$. Then, we can update ${\bf U}=1 \to 2 \to 3$, $\mathcal{I}_{+}=\{1,2,3\}$ and $\mathcal{I}_-=\O$. Finally, we obtain the feasible path ${\cal U}^\star={\bf U}$ as the output of Algorithm 2, as in line 8. 

\subsubsection{Algorithm 3} \textit{Dynamic programming}

Although having a lower complexity, Algorithm 2 obtains a much degraded performance compared with Algorithm 1. This motivates us to propose Algorithm 3, which is based on DP and takes into account the future outcome when selecting a path. It will be shown later that the DP-based algorithm approaches the optimal solution with a considerably reduced complexity. 

Denote $G=(\cal{K},A),$ where $\cal{K}$ is the set of GUs and ${\cal A}=\{a_{jk}\}$ is the set of the summation of travel time from GU $j \to k$ and the data transmission time to GU $k$, e.g., $a_{jk}=l_{j \to k}/V_{\max}+\tau_k, j \ne k$. In this work, since we do not consider $a_{jk}$ with $j = k$, thus, $a_{jk}$  will henceforth be referred to as $a_{jk}$ with $j \ne k$. Moreover, $a_{jk}$ is feasible if it satisfies the RT constraint, i.e., $a_{jk} \le \eta_k$. As in the first step of Algorithms 1 and 2, we check the feasibility by considering the RT constraint from UAV station to GU $k$, as in lines 3 of Algorithm 3. Concretely, if there exists a value of $a_{0k}$ which does not satisfy the RT constraint, a feasible path will not exist. Associate with each GU $k \in \mathcal{K}$ a time window $[0,\eta_k]$ and a data transmission time $\tau_k.$ 

A state $(\mathcal{S},k)$ is defined as: $\mathcal{S}$ is an unordered set of visited GUs, $k$ is the last visited GU in $\mathcal{S}$. Define $C(\mathcal{S},k)$ as the least cost (e.g., summation of traveling time and data transmission time to GU $k$) of path starting at UAV station, passing through each GU of $\cal{S} \subset \mathcal{K}$ exactly once, ending at GU $k$. Without loss of generality, we initialize the cost function $C$ as $C(\{0\},0)$ equals to zero, whereas the first and second elements represent for the UAV station. The $C(\mathcal{S},k)$ is calculated by solving the following equation \cite{Dumas}
\begin{align}
        \label{DP:cost}
        \begin{array}{l}
        C(\mathcal{S},k) = \mathop {\min }\limits_{\left( a_{jk} \right) \in {\cal A}} \{ C(\mathcal{S}\backslash \{ k\} ,j) + C(\{j,k\},k)|  \\ \qquad \qquad \qquad \qquad 
         C(\mathcal{S}\backslash \{ k\} ,j)+ C(\{j,k\},k) \le \eta_k\}. 
        \end{array}
\end{align}
where $C(\{j,k\},k)= a_{jk} = l_{j \to k}/V_{\max} +\tau_k$, $\mathcal{S} \subset \mathcal{K}, j \; \text{and} \; k \in \mathcal{S}$. 

We denote ${\cal B}_m^k$ with $k \in {\cal K}, m=1,2,\dots,K+1$ as the set containing the last visited GU before visiting GU $k$ in step $m$, as in lines 2 and 7. Specifically, when an UAV starts from ground station, there is no visited GU before this, i.e., ${\cal B}_1^0=\O$ as in line 2. Let $\Xi_m$ denote the set of all feasible states $(\mathcal{S},k)$, where $|\mathcal{S}|=m$. In order to obtain $\Xi_m$ from $\Xi_{m-1}$, we do following steps. For each state $(\mathcal{S},j)\in \Xi_{m-1},$ we consider a new state $((\mathcal{S},j) \cup k,k)$, lines 5 to 7. This state can be added to $\Xi_m$ iff it satisfies the RT constraint and is not yet stored, as in line 6. In the case that this state is already stored in $\Xi_m$, we only keep the state having minimum cost of $C((\mathcal{S},j) \cup k,k)$, as in line 6. Let assume that there exist two states with the corresponding cost functions $C_1(\mathcal{S},k)$ and $C_2(\mathcal{S},k)$, respectively. If $C_1(\mathcal{S},k)<C_2(\mathcal{S},k)$, then, the second state will be eliminated. The goal of DP algorithm is to take all the feasible paths satisfying constraints C1 and C2. Since we only store the state with lowest value of $C(\mathcal{S},k), k=1,\dots,K$ for each state $({\cal S},k)$. Thus, at the end of Algorithm 3, when $|{\cal S}|=K+1$, we can achieve maximally $K$ states $({\cal S},k) \in \Xi_{K+1}$. For each state $({\cal S},k)$, the visiting order ${\bf u}^\star$ is obtained by checking for backward from $\Xi_{K+1}$ to $\Xi_1$ based on ${\cal B}_m^k$, as in line 10. Finally, feasible Hamiltonian cycle paths ${\cal U}^\star$ with $|{\cal U}^\star|\le K$ is acquired at the output of DP algorithm, as in line 12.

To make it easy to understand, the DP-based algorithm is illustrated in Tables \ref{table21} and \ref{table3}. More specifically, we consider ${\cal K}=\{1,2,3\}$, ${\boldsymbol \eta} = \{2,2,4\}$ seconds, $0$ denotes the UAV station, Table \ref{table21} is the travel time between GUs. Due to the RT constraint and the condition of storing one state with minimum cost $C({\cal S},k)$, we cannot keep all states into consideration. For example, when ${\cal S}=3$, we only achieve one final state, i.e., $(\{0,1,2,3\},3)$. For the last state $ \{0,1,2,3\} \in \Xi_4$, we can check for backward from $\Xi_4$ to $\Xi_1$ to find the feasible path ${\bf u}^\star$. More specifically, from state $ \Xi_4$, we can find out that $1 \in {\cal B}_4^3$ is the last visited GU before visit 3. Next, by considering the state $({\cal S},1)\in \Xi_3$, $2 \in {\cal B}_3^1$ is the last visited GU before visit 1. Similarly, we check for backward until reaching UAV station $0$. Finally, the visiting order ${\bf u}^\star$ is obtained, i.e., ${\bf u}^\star = \{ 0 \to 2 \to 1 \to 3 \to 0\}$. The complexity of the DP-based algorithm is $\mathcal{O}(K^2\times2^K)$ in the worst case \cite{Held}. Moreover, details of this method are described in Algorithm 3.

\begin{table}[t]
	\caption{Illustration for the value of ${\cal A}=\{a_{jk}, j, k \in {\cal K}, \}$ in DP Algorithm}
	\label{table21}
	\centering
	\small
	\begin{tabular}{|p{1cm}|p{1cm}|p{1cm}|p{1cm}|p{1cm}|}
		\hline
		GU & 0 & 1 & 2 &  3\\
		\hline
		0 &$+\infty$& 1 & 1.4 & 1.2 \\
		\hline
		1 & $1$& $+\infty$& 0.5 & 1.5 \\
		\hline
		2 & 1.4& 0.5 & $+\infty$ & 2 \\
		\hline
		3 & 1.2& 1.5 & 2 & $+\infty$ \\
		\hline
	\end{tabular}
\end{table}
\begin{table}[t]
	\caption{Illustration for DP Algorithm}
	\label{table3}
	\centering
	\small
	\begin{tabular}{|p{0.5cm}|p{1.5cm}|p{1cm}|p{1.5cm}|p{1cm}|}
		\hline
		$\Xi$ & $\cal S$ & $k$ & ${\cal B}_m^k$ & $C({\cal S},k)$\\
		\hline
		$\Xi_1$ & $\{0\}$& $\O$ & ${\cal B}_1^0=\O$ & $0$ \\
		\hline
		$\Xi_2$ & $\{0,1\}$& $1$ & ${\cal B}_2^1=\{0\}$ & 1 \\
		\hline
		$\Xi_2$ & $\{0,2\}$& $2$ & ${\cal B}_2^2=\{0\}$ & 1.4 \\
		\hline
		$\Xi_2$ & $\{0,3\}$& $3$ & ${\cal B}_2^3=\{0\}$ & 1.2 \\
		\hline
		$\Xi_3$ & $\{0,1,2\}$& $2$ & ${\cal B}_3^2=\{1\}$ & 1.5 \\
		\hline
		$\Xi_3$ & $\{0,1,2\}$& $1$ & ${\cal B}_3^1=\{2\}$ & 1.9 \\
		\hline
		$\Xi_3$ & $\{0,1,3\}$& $3$ & ${\cal B}_3^1=\{1\}$ & 2.5 \\
		\hline
		$\Xi_4$ & $\{0,1,2,3\}$& $3$ & ${\cal B}_4^3=\{1\}$ & 3.4 \\ 
		\hline
	\end{tabular}
\end{table}
%%%%%%%%%%%%%%%%%%%%%%%%%%%%%%%%%%%%%%%%%%%%%%%%
\subsection{Minimization of the UAV's Energy Consumption with given path}
\label{Subsec:B}
The previous section designs the paths based on the UAV maximum speed. While this method is preferred to minimize the traveling time, it might not be energy-efficient since it over-estimates the RT constraints. In this section, we minimize total energy consumption of the UAV via the joint optimization of UAV velocities over each given path in the feasible set $\mathcal{U}^\star$, e.g., the output of Algorithms~\ref{alg:1}, \ref{alg:2}, and \ref{alg:3}. The energy minimization problem is formulated as
% EQ8
\begin{align}
\label{OP:E}
\mathcal{P}_2: &\mathop {\min }\limits_{\{v_i\}_{i=1}^{K+1}}~~ {\sum}_{i = 1}^{K+1} ({E_{{\rm fly},i}(v_i) + E_{{\rm hov},i} + E_{{\rm com},i}})  \\
\mathtt{s.t.}~~~
&\text{C1}:~\sum_{i=1}^{k} \left(\frac{d_i}{v_i} + \tau_{u_i} \right) \le \eta_{u_k}, 1\le k \le K \notag\\
&\text{C2}:~0 \le v_i \leq V_{\rm max}, \;\;\; i=1,\dots,K+1. \notag \\
&\text{C3}: |v_{i+1} - v_{i}| \le \Delta V, \forall 1 \le i \le K,
\end{align}
%\mathcal{P}_1
%%%%%%%%%%%%%%%%%%%%%%%%%%%%%%%%%%%%%%%%%%%%%%%%

Because $E_{{\rm com},i}$ and $E_{{\rm hov},i}$ do not depend on $v_i$, they can be removed from the objective function of \eqref{OP:E} without loss of generality. Since function $\frac{1}{x}$ is convex in ${\{v_i\}_{i=1}^{K+1}} \in \mathbb{R}^+$, constraint C1 in \eqref{OP:E} is convex. The most challenging is the term $E_{{\rm fly},i}(v_i)$. 
\begin{lemma}
        \label{lemma:2}
        The energy consumption  $E_{{\rm fly},i}(v_i)$ is convex.
\end{lemma}
\begin{IEEEproof}
        From \eqref{eq:5} and \eqref{eq:6} we have
        \begin{align}
        \label{AppI_0}
        E_{{\rm fly},i}(v_i) = P_0 d_i \left(\frac{1}{v_i} + \alpha_1 v_i\right) +& P_1 d_i \sqrt{\sqrt{v_i^{-4} + \alpha_2^2}-\alpha_2}\notag \\
        +& \alpha_3 d_i v_i^2,
        \end{align}        
        The second derivative of $E_{{\rm fly},i}(v_i)$, after some manipulations, can be expressed as
        \begin{align}
        \label{AppI_1}
        &\frac{d^2}{dv_i^2}E_{{\rm fly},i}(v_i) = \frac{2P_0d_i}{v_i^3} + 2\alpha_3d_i + P_1d_i \beta,
        \end{align}
        where
        \begin{small}
                \begin{align}
                \label{AppI_2}
                \beta  &= \frac{1}{{{v_i^6}\sqrt {\alpha _2^2 + {v_i^{ - 4}}} \sqrt {\sqrt {\alpha _2^2 + {v_i^{ - 4}}}  - {\alpha _2}} }} \times \nonumber\\
                &\Big( 5 - \frac{2}{1 + \alpha_2^2 v_i^4} - \frac{1}{\underbrace{\alpha_2^2 v_i^4 + 1  - \alpha_2 v_i^2 \sqrt{\alpha_2^2 v_i^4 + 1}}_{\beta_1}} \Big).
                \end{align}
        \end{small}
        
        Denote $X = \alpha_2 v_i^2 \ge 0$, then we can express $\beta_1 = \alpha_2^2 v_i^4 + 1  - \alpha_2 v_i^2 \sqrt{\alpha_2^2v_i^4 + 1} = X^2 + 1 - X\sqrt {{X^2} + 1}$. Since
        %       \begin{align}
        $X\sqrt {{X^2} + 1}  \le \frac{{2{X^2} + 1}}{2}$,
        %       \end{align}
        %
        it yields
        \begin{align}
        \label{AppI_4}
        {\beta_1} \ge {X^2} + 1 - \frac{{2{X^2} + 1}}{2} = \frac{1}{2}. 
        \end{align}

        In addition, since $1 + \alpha_2^2v_i^4 \ge 1$, we obtain the term in bracket in \eqref{AppI_2} is always greater than or equal to 1. Thus, $\beta > 0, \forall v_i$. Since $P_0, d_i, \alpha_3$ are also positive, from \eqref{AppI_1} we conclude that the second derivative of $E_{{\rm fly},i}(v_i)$ is always positive, which proves the convexity of $E_{{\rm fly},i}(v_i)$. 
\end{IEEEproof}

By using Lemma~\ref{lemma:2}, we observe that problem $\mathcal{P}_2$ is convex since the objective and all constraints are convex. Thus, it can be solved by using the standard methods \cite{Boyd}.

 %%%%%%%%%%%%%%%%%%%%%%%%%%%%%%%%%%%%%%%%%%%%%%%%%%%%%%%
%\begin{table}[t]
%	\caption{System Setup for Numerical Simulations}
%	\label{table5}
%	\centering
%	\small
%	\begin{tabular}{|p{5.5cm}|p{2.7cm}|}
%		\hline
%		\textbf{Parameters} & \textbf{Values}\\
%		\hline
%		UAV altitude & H = 50 meters\\
%		\hline
%		Maximum UAV speed & $V_{\rm max}$ = 50 m/s \\
%		\hline
%		Minimum requested timeout value & $\eta_{\min }=5$ seconds\\
%		\hline
%		Maximum requested timeout value & $\eta_{\rm max }=60$ seconds\\
%		\hline
%		Communication related power consumption of UAV & $P_{com}= 5 $ W\\
%		\hline
%		The communication Bandwith & $B=$ 2 MHz\\
%		\hline
%		Path loss exponent & \bl{$\alpha=2.3$} \\
%		\hline
%		\bl{Rician factor $G$} & \bl{15} dB\\
%		\hline
%		\bl{Reference channel power gain $\mu_0$} & \bl{-50} dB\\
%		\hline
%		Noise power & \bl{$\sigma^2=$-110 dBm}\\
%		\hline
%		Total number of iteration & 1000\\
%		\hline
%		Packet size & $Q_k=5$ Mbits\\
%		\hline
%		UAV's coverage area & 400 m x 400 m\\
%		\hline
%		UAV ground station's location & (1.5m, 398m)\\
%		\hline
%	\end{tabular}
%\end{table}

%%%%%%%%%%%%%%%%%%%%%%%%%%%%%%%%%%%%%%%%%%%%%%%%%%%%
% FIG2 FIG2 FIG2 FIG2 FIG2 FIG2 
\begin{figure*}[t]
	\begin{subfigure}{0.5\textwidth}
		\centering
		\includegraphics[width=9cm,height=7cm]{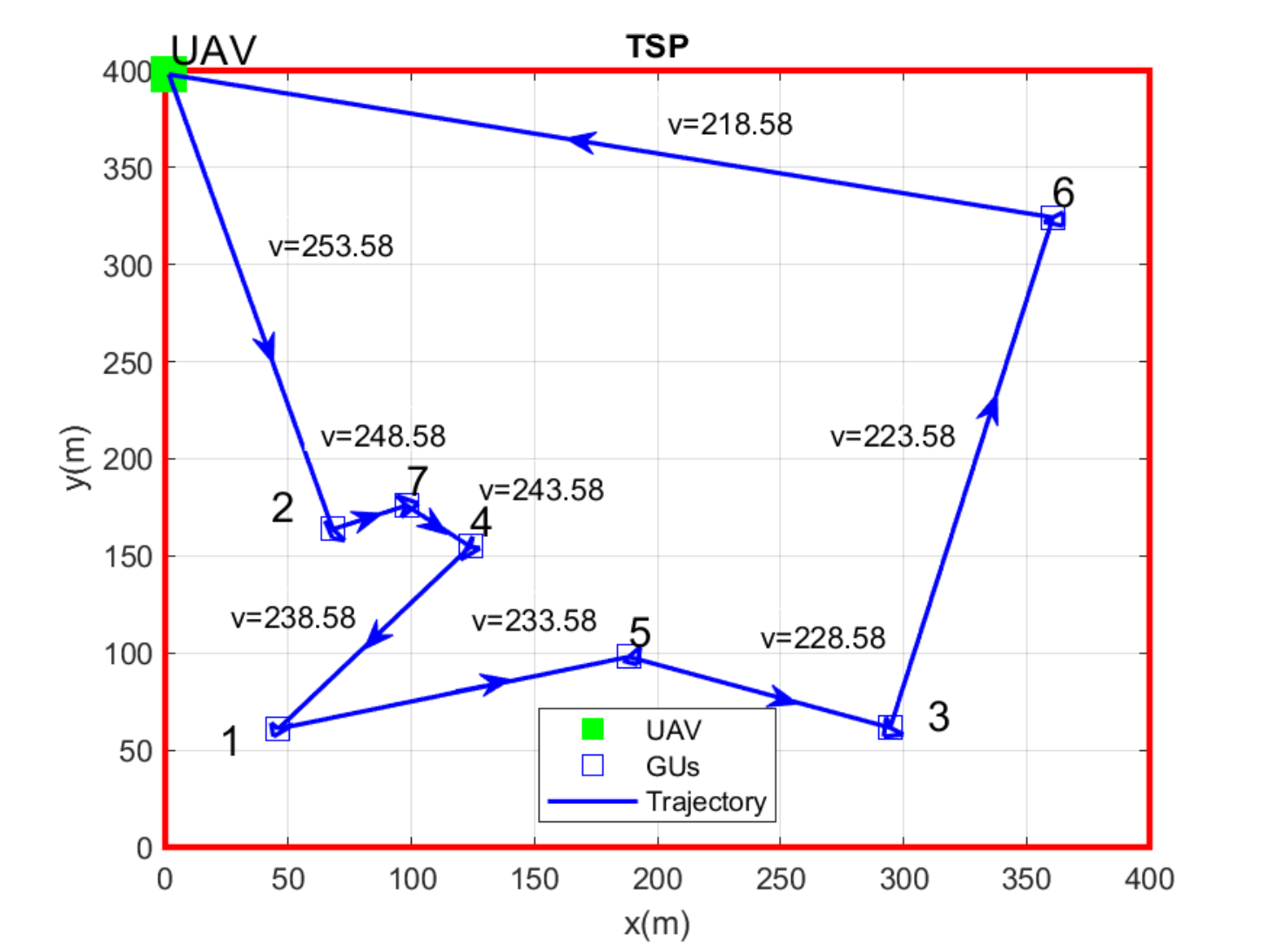}
		\caption{Traditional TSP, $\sum\limits_{i = 1}^{K+1} {E_i}= 6175 \;Joules$.}
		\label{fig:2a}  
	\end{subfigure}
	\begin{subfigure}{0.5\textwidth}
		\centering
		\includegraphics[width=9cm,height=7cm]{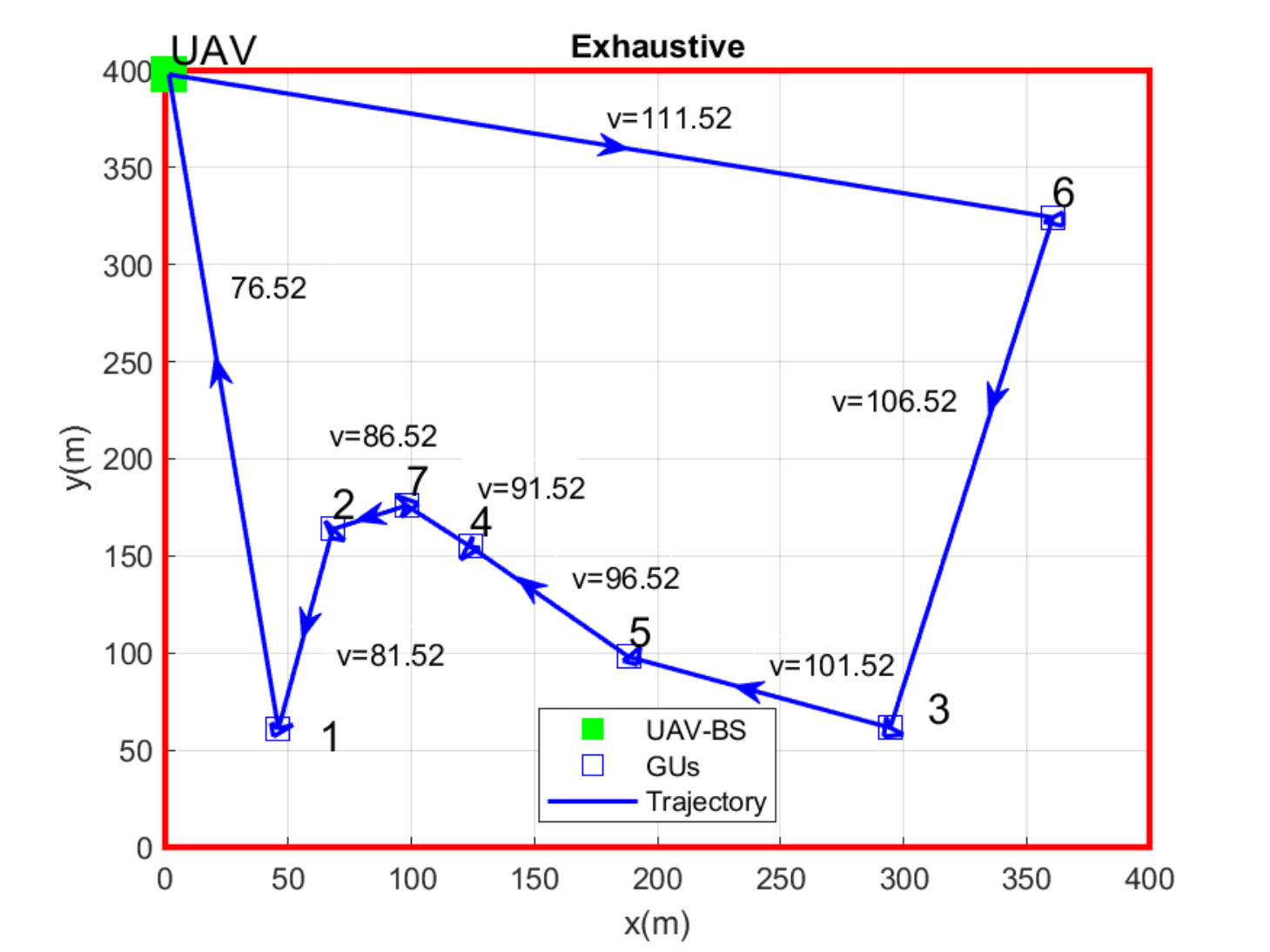}
		\caption{Exhaustive search, $\sum\limits_{i = 1}^{K+1} {E_i}= 3944 \;Joules$.}
		\label{fig:2b}  
	\end{subfigure}
	\begin{subfigure}{0.5\textwidth}
		\centering
		\includegraphics[width=9cm,height=7cm]{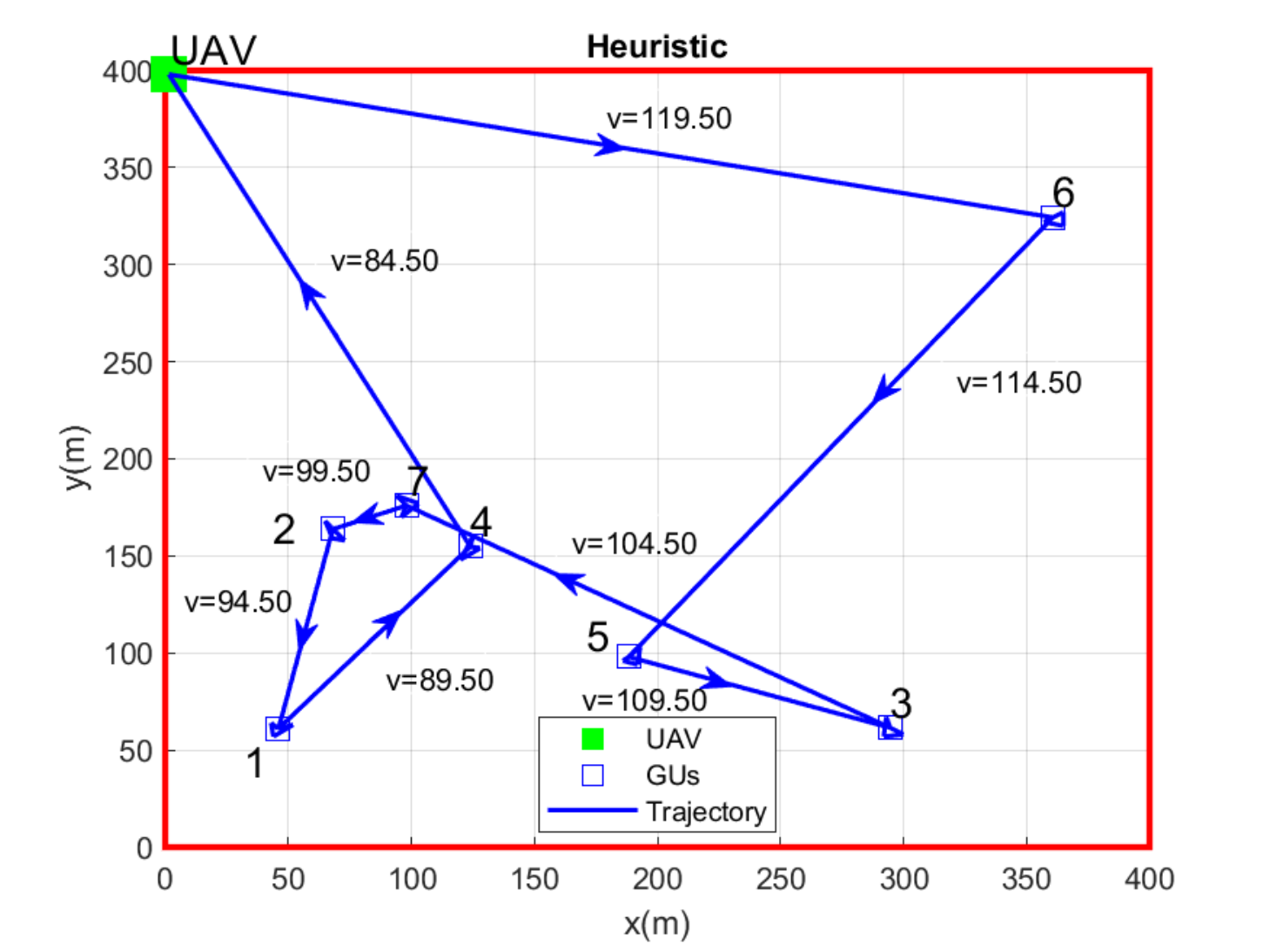}
		\caption{Heuristic, $\sum\limits_{i = 1}^{K+1} {E_i}= 4484 \;Joules$.}
		\label{fig:2c}  
	\end{subfigure}
	\begin{subfigure}{0.5\textwidth}
		\centering
		\includegraphics[width=9cm,height=7cm]{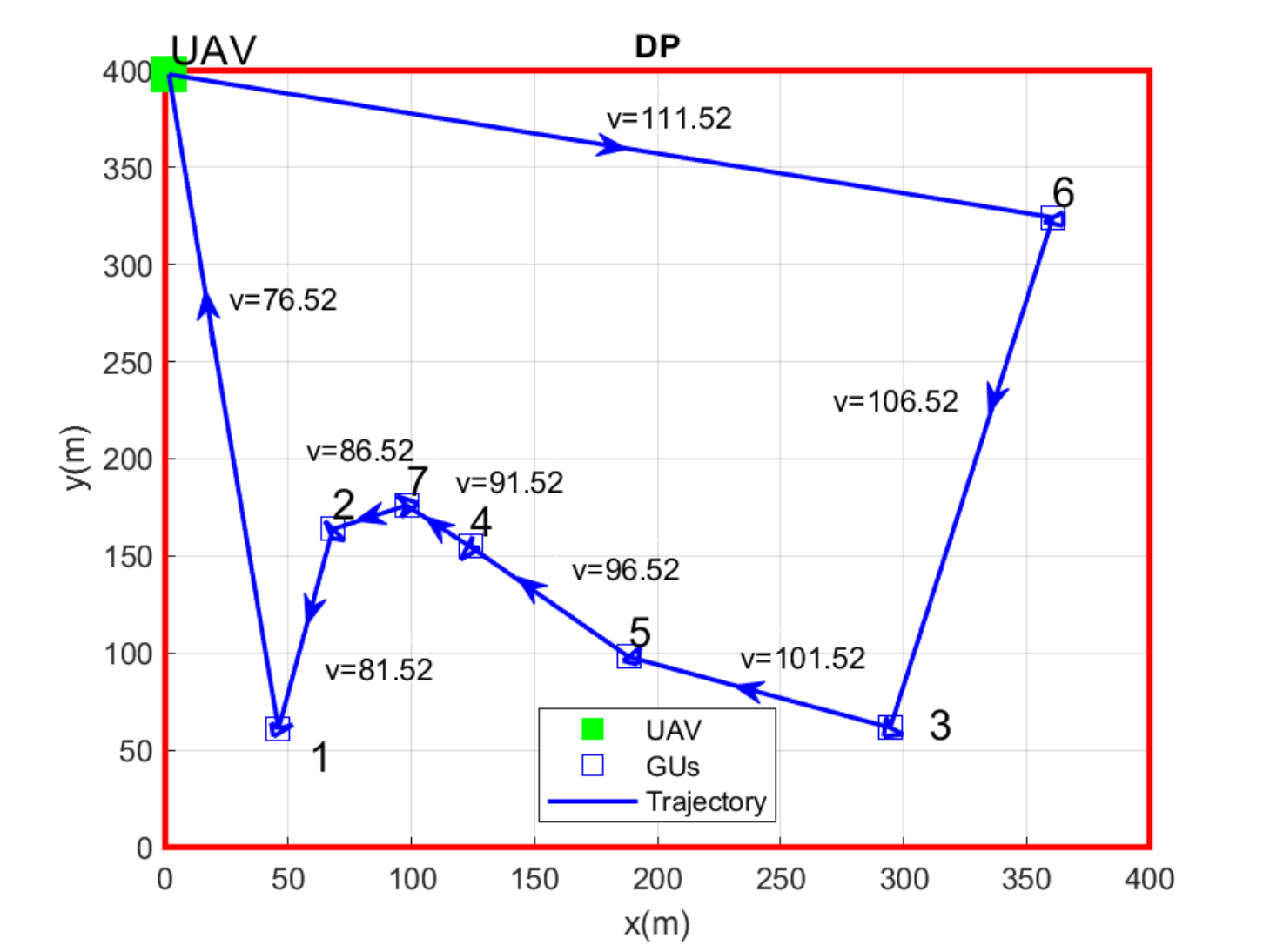}
		\caption{Dynamic programming, $\sum\limits_{i = 1}^{K+1} {E_i}= 3944 \;Joules$.}
		\label{fig:2d}  
	\end{subfigure}   
	\caption{ Comparison of UAV's trajectories with different path designs  }
	\label{fig:2}
\end{figure*}
% FIG2 FIG2 FIG2 FIG2 FIG2 FIG2
%%%%%%%%%%%%%%%%%%%%%%%%%%%%%%%%%%%%%%%%%%
%%%%%%%%%%%%%%%%%%%%%%%%%%%%%%%%%%%%%%%%%%%%%%%%
% FIG3 FIG3 FIG3 FIG3 FIG3
\begin{figure}[t]
        \centering      
        \includegraphics[width=9cm,height=7cm]{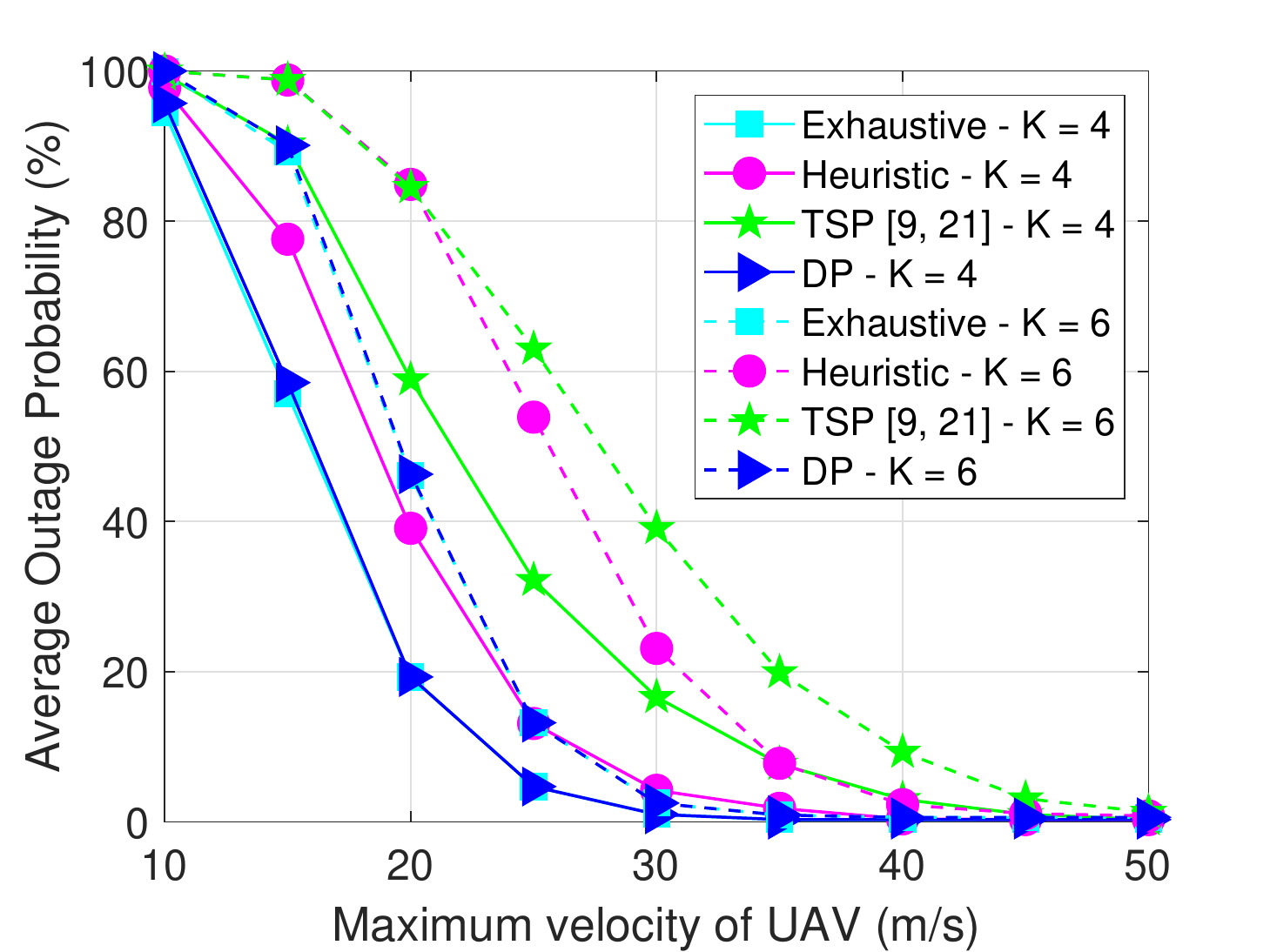}
        \caption{ Average OP ($\% $) versus $V_{\rm max}$ (m/s)}
        \label{fig:3}
\end{figure}
% FIG3 FIG3 FIG3 FIG3 FIG3
%%%%%%%%%%%%%%%%%%%%%%%%%%%%%%%%%%%%%%%%%%%%%%%%
% FIG4 FIG4 FIG4 FIG4 FIG4 
\begin{figure}[t]
        \centering      
        \includegraphics[width=9cm,height=7cm]{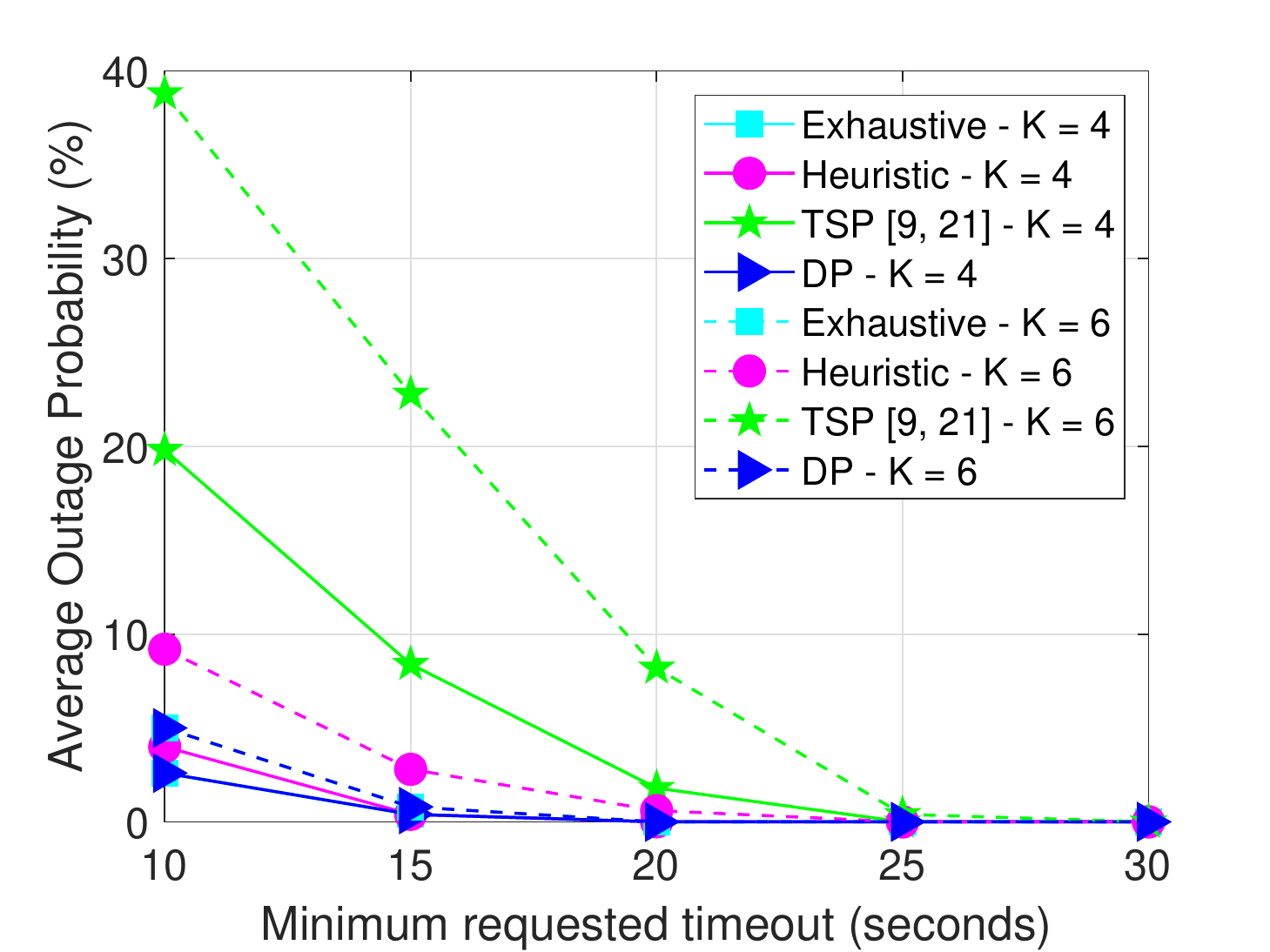}
        \caption{ Average OP ($\% $) versus minimum requested timeout}
        \label{fig:4}
\end{figure}
% FIG4 Barcelona 5-0 LeganesIG4 FIG4 FIG4 FIG4
%%%%%%%%%%%%%%%%%%%%%%%%%%%%%%%%%%%%%%%%%%%%%%%%
% FIG5 FIG5 FIG5 FIG5 FIG5 
\begin{figure}[t]
	\centering      
	\includegraphics[width=9cm,height=7cm]{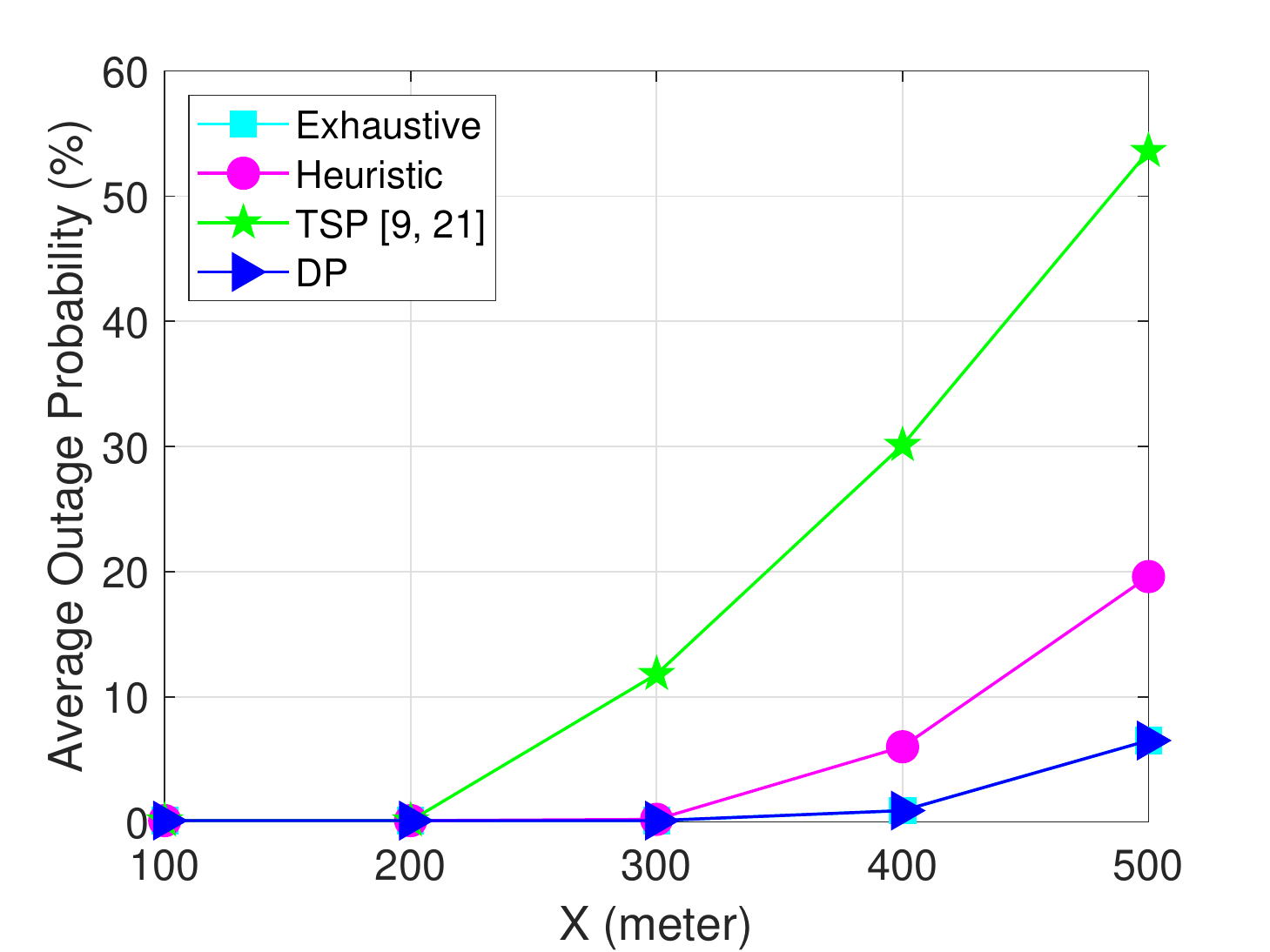}
	\caption{ Average OP versus network size, e.g., $A=x^2$ $(m^2)$}
	\label{fig:5}
\end{figure}
% FIG5 FIG5 FIG5 FIG5 FIG5
%%%%%%%%%%%%%%%%%%%%%%%%%%%%%%%%%%%%%%%%%%%%%%%%
% FIG6 FIG6 FIG6 FIG6 FIG6
\begin{figure}[t]
	\centering      
	\includegraphics[width=9cm,height=7cm]{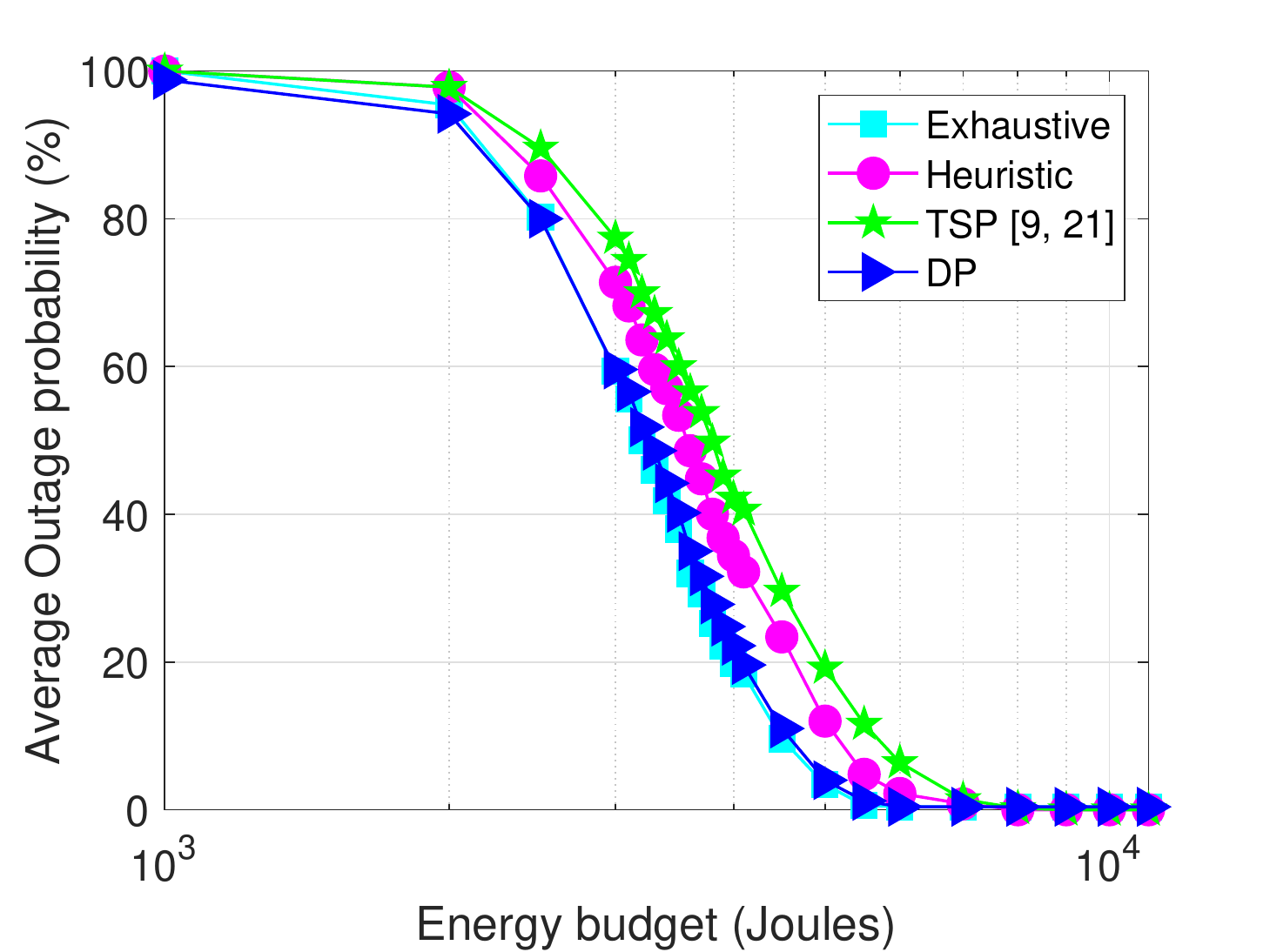}
	\caption{ Average OP ($\% $) versus energy budget}
	\label{fig:6}
\end{figure}
% FIG6 FIG6 FIG6 FIG6 FIG6 
\section{Simulation Results}
\label{Sec:Num}

This section provides numerical results to validate the proposed designs. The parameters are set as follows: $H=50$ meters,  $B=2$ MHz, path loss exponent $\alpha=2.3$, $\sigma^2=-110$ dBm, $P_{com}=5$ W, Rician factor G = 15 dB, UAV's coverage area is 400 m x 400 m, UAV ground station is located at (1.5m, 398m). On a more general level, we perform 1000 independent trials of Monte-Carlo simulations. In details, for each iteration, we deploy a random GUs topology distributed in the considered area, the RT constraints are uniformly ranging between $\eta_{\min }$ and $\eta_{\max }$. Moreover, the channel coefficient $g_k$ is also regenerated for each iteration. The proposed solutions are compared with a solution in \cite{Y_Zeng_1,Y_Zeng}, which is based on the TSP. Specifically, since the heuristic and DP algorithms can find maximally $1$ and $\Theta \le K$ feasible paths, respectively. In order to guarantee that the exhaustive method is always an upper bound, this algorithm takes $\Psi$ ($\Psi\ge \Theta$) shortest feasible paths while the heuristic and DP algorithms take all feasible paths into consideration. %Unless stated, other parameters in the simulations are given in Table \ref{table5}. 

In Fig.~\ref{fig:2}, we illustrate the difference trajectory designs, i.e., TSP, exhaustive search, heuristic, and DP, with $\epsilon=0.001$, $Q_k$ = 50 Mbits, $\eta_{\min}=5$ seconds, $\eta_{\max }=17$ seconds, $K=7$, $E_{\rm tot}=100 $ KJoules. For the purpose of a fair comparison, $V_{\max}$ is assumed to be sufficiently large so that all methods exists one path complying with the latency constraints. The arrows in Fig.~\ref{fig:2} denote the moving direction of UAV. While the TSP always follows the shortest path and does not take latency into account, the others select the path with minimum energy consumption via optimizing the traveling velocities. Therefore, the energy consumption of TSP is higher than others. Moreover, the heuristic method selects the closet GU having minimum RT value as the next visited GU, it leads to the longer traveling distance. This explains why the energy consumption of heuristic is higher than exhaustive and DP methods. Specifically, the DP method may obtain the same trajectory design compared to that of exhaustive search. It shows the superiority of this scheme compared to other ones. %We will discuss about each method in more detail on the following figures.

Next, we evaluate the proposed trajectory designs via the outage probability metric (OP). Moreover, the in-feasibility also occurs if all the paths, which is obtained from Algorithm 1 (or Algorithms 2, 3), do not satisfy the energy budget constraint, i.e., C3 in \eqref{OP:energy}. Fig.~\ref{fig:3} presents the OP of the proposed algorithms and the reference as a function of $V_{\rm max}$ with the RT requirements $\eta_k$ ranging between 22 and 60 seconds, the energy budget $E_{\rm tot}=500$ KJoules, $B=3$ MHz, $\epsilon=0.001$, $G=$ 15 dB. It is shown that the proposed algorithms significantly improve the OP compared with the reference for all values of $V_{\rm max}$. Specifically, at $V_{\rm max} = 40$ m/s and $K=6$, the exhaustive search and dynamic programing algorithms can find the trajectory that satisfies all the GUs' RT constraints with high probability and the heuristic algorithm achieves less than 1.7\% OP. Whereas the reference scheme imposes 9.3\% OP. The OP of all schemes can be reduced by increasing $V_{\rm max}$, which is because a higher $V_{\rm max}$ results in a lower traveling time between the GUs. Consequently, it is highly probable for the UAV to satisfy the GUs' RT.

%%%%%%%%%%%%%%%%%%%%%%%%%%%%%%%%%%%%%%%%%%%%%%%%
% FIG7 FIG7 FIG7 FIG7 FIG7 
\begin{figure}[t]
	\centering      
	\includegraphics[width=9cm,height=7cm]{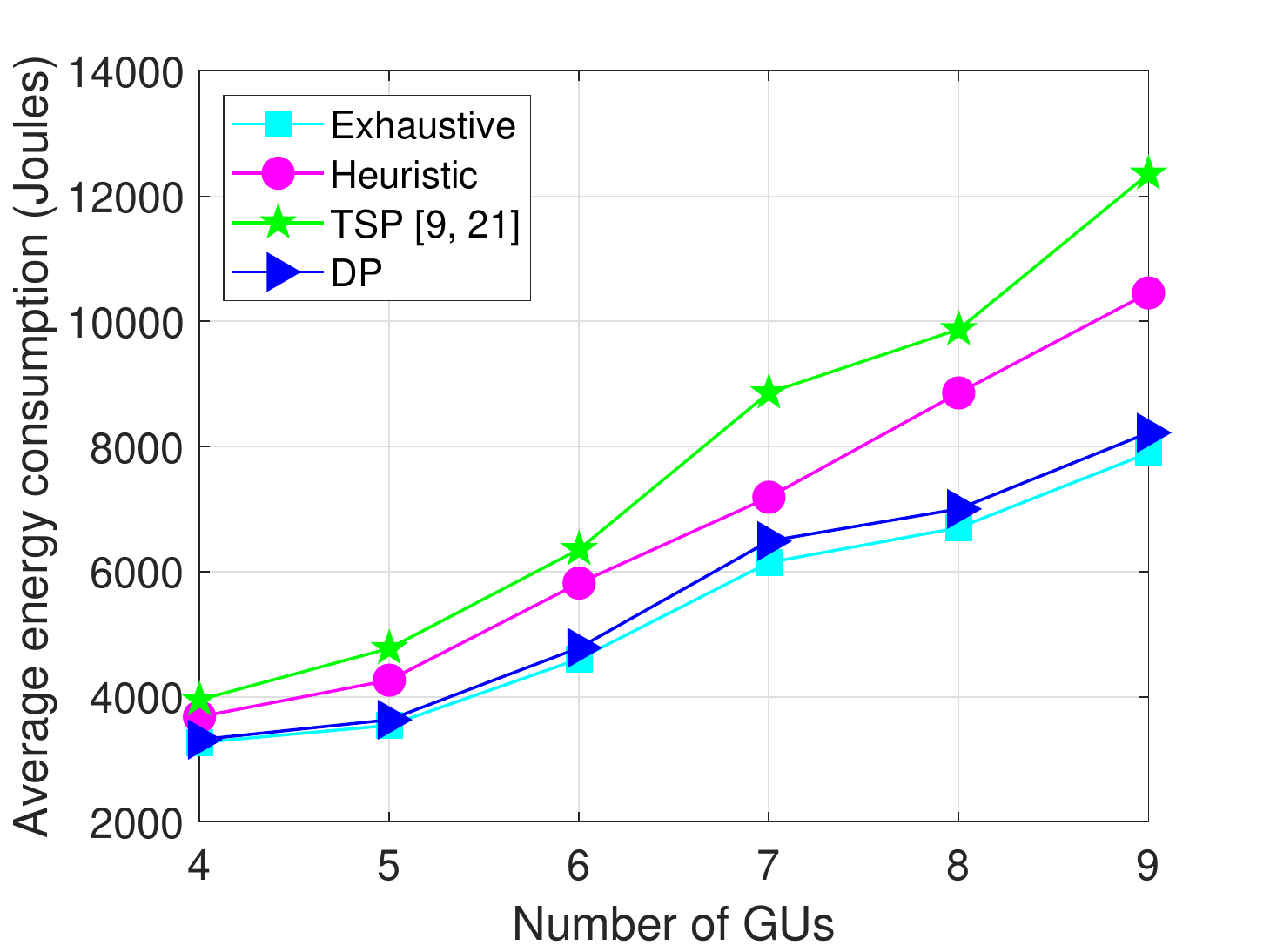}
	\caption{ Average energy consumption vs. number of GUs.}
	\label{fig:7}
\end{figure}
% FIG7 FIG7 FIG7 FIG7 FIG7 
%%%%%%%%%%%%%%%%%%%%%%%%%%%%%%%%%%%%%%%%%%%%%%%%
\begin{figure}[t]
	\centering      
	\includegraphics[width=9cm,height=7cm]{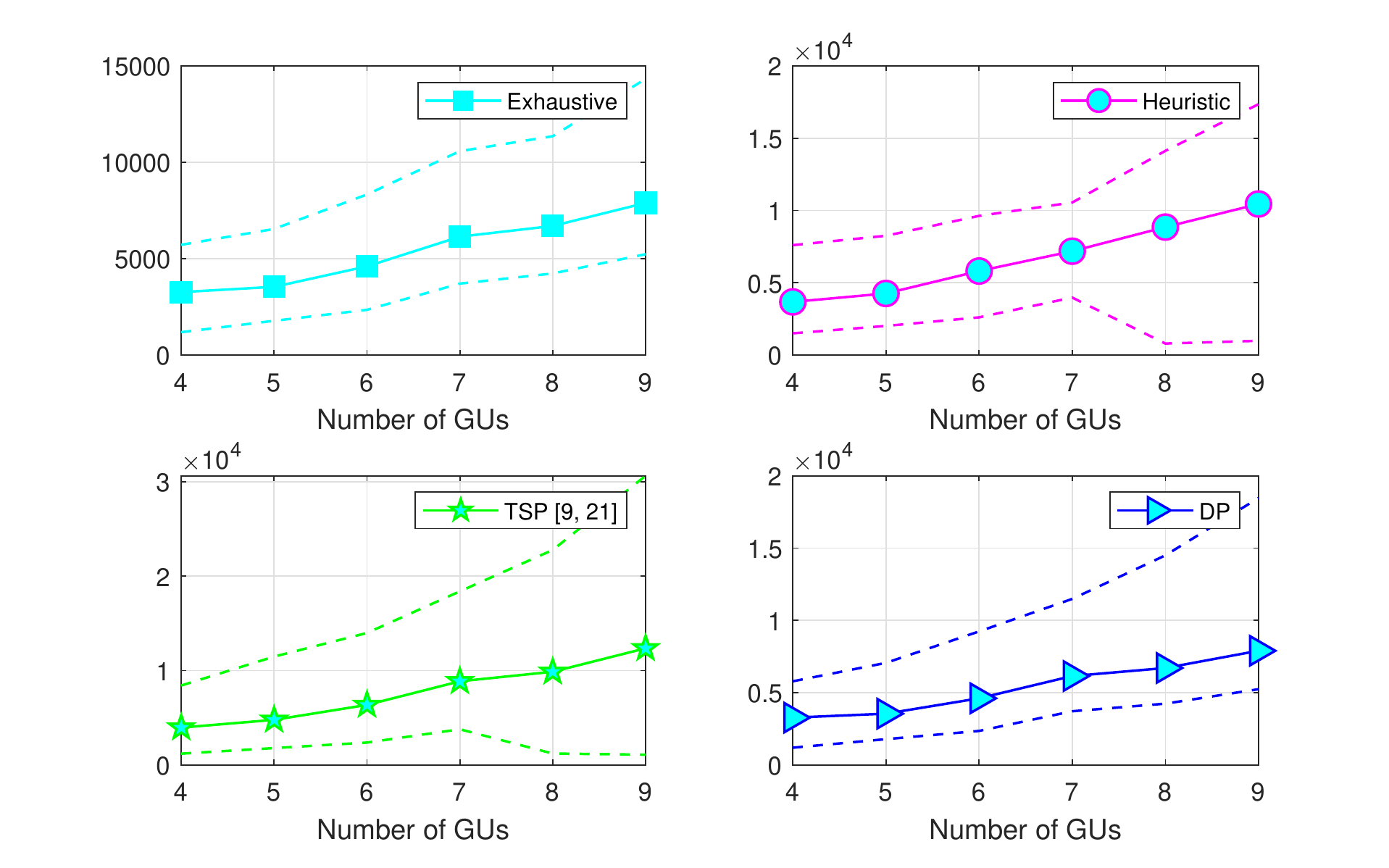}
	\caption{ Average, minimum, and maximum values of energy consumption (Joules).}
	\label{fig:8}
\end{figure}
%%%%%%%%%%%%%%%%%%%%%%%%%%%%%%%%%%%%%%%%%%%%%%%%
\begin{figure}[t]
	\centering      
	\includegraphics[width=9cm,height=7cm]{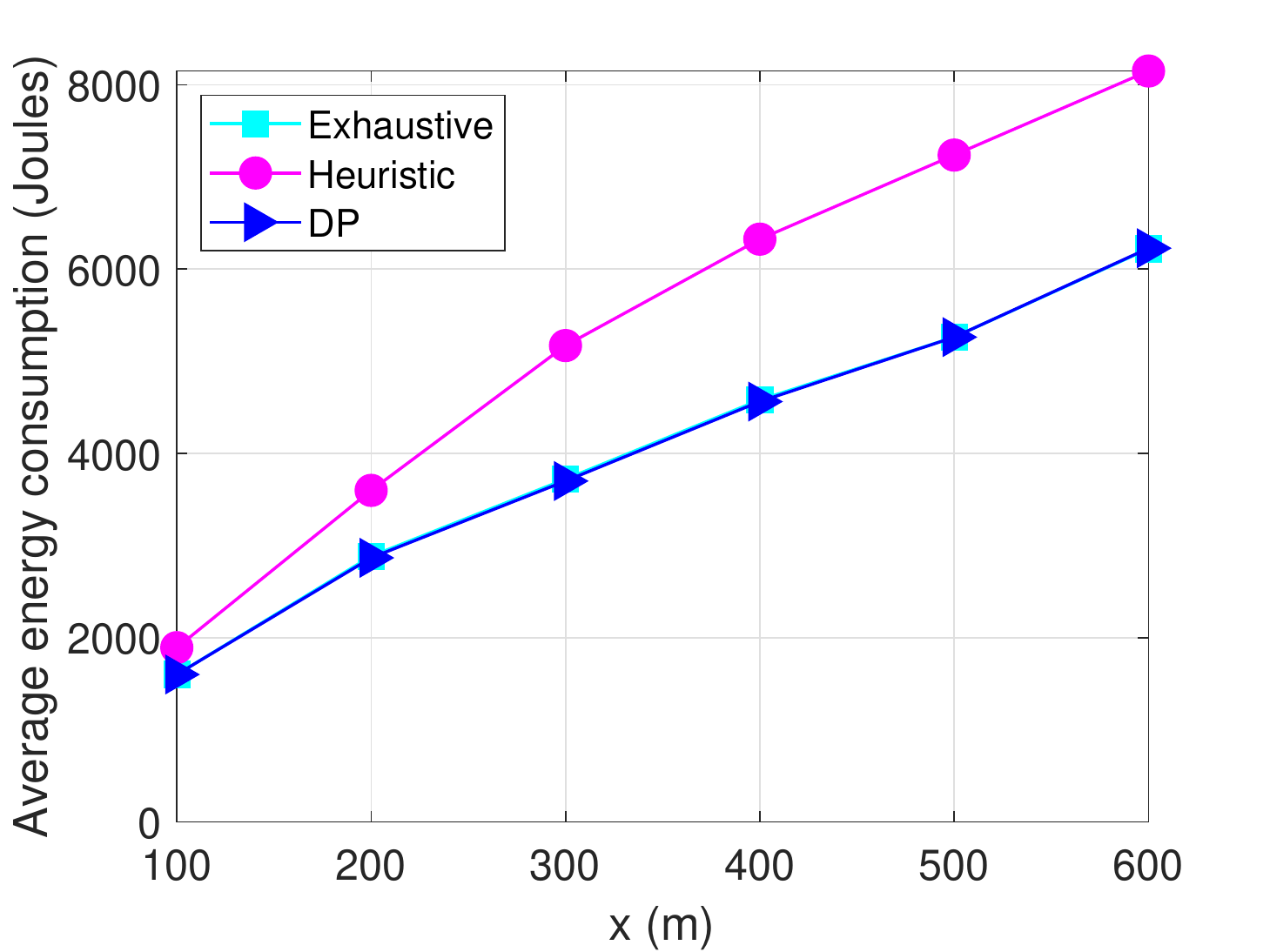}
	\caption{ Average energy consumption vs. network size}
	\label{fig:9}
\end{figure}
%%%%%%%%%%%%%%%%%%%%%%%%%%%%%%%%%%%%%%%%%%%%%%%%
\begin{figure}[t]
	\centering      
	\includegraphics[width=9cm,height=7cm]{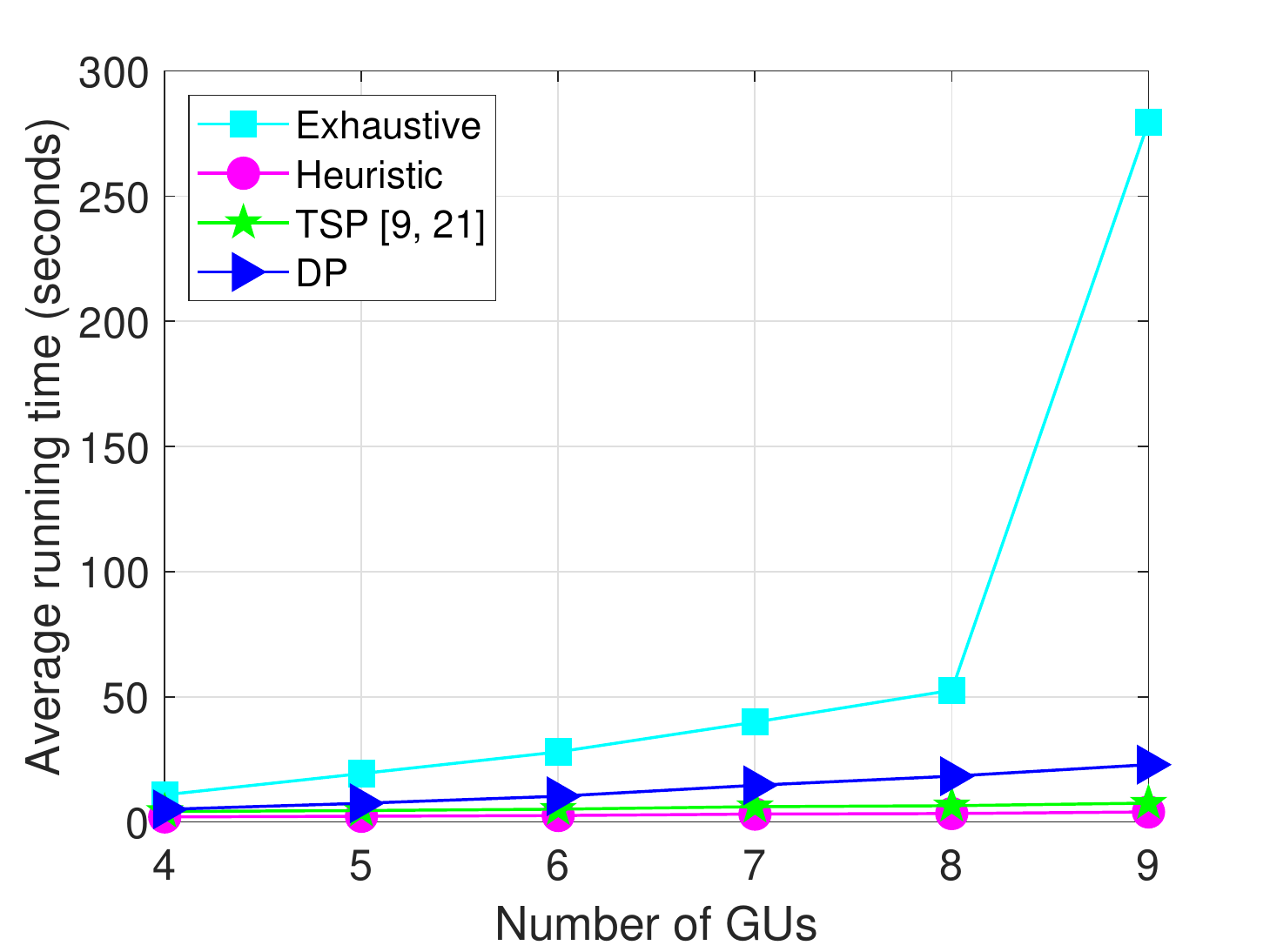}
	\caption{ Average calculation time vs. number of GUs.}
	\label{fig:10}
\end{figure}
%%%%%%%%%%%%%%%%%%%%%%%%%%%%%%%%%%%%%%%%%%%%%%%%

In Fig. \ref{fig:4}, the OP is presented as a function of minimum RT value $\eta_{\min }$ (seconds), while $\eta_{\max } = 65$ seconds, $V_{\rm max}=45$ m/s, and $E_{\rm tot}=500$ KJoules, $Q_k=10$ Mbits, $B=3$ Mhz, $G=$ 15 dB. Similar to Fig. \ref{fig:3}, the DP achieves almost the same outage performance as the exhaustive search while it significantly outperforms the heuristic and reference algorithms. Specifically, at $\eta_{\min }=15$ seconds and $K=6$, the OP values of both exhaustive search and dynamic programing algorithms equal to 5\% and the heuristic-based algorithm achieves less than 4.2 \% OP. Whereas the reference scheme imposes 38.8\% OP. It is found that at a lower value of $\eta_{\min }$, the outage performance is degraded. This is expected since allocating more speed is needed to satisfy the GUs' RT, but the $V_{\max}$ is limited. Furthermore, in Figs. \ref{fig:3} and \ref{fig:4}, the average outage probability decreases when we decrease the number of GUs from 6 to 4.

Fig. \ref{fig:5} represents the OP as a function of network size (i.e., $x$ (meters)) with $K=7,$ $V_{\rm max}=45$ m/s, $\eta_{\min}=15$ seconds, $\eta_{\max}=65$ seconds, $Q_k=$ 10 Mbits, $G=$ 15 dB, $B=3$ Mhz and $E_{\rm tot}=500 $ KJoules. Whereas the UAV's coverage area is assumed to be a square and it can be calculated as $A=x^2$ $(m^2)$, e.g., Fig. \ref{fig:1}. It is observed that with the increasing of $x$, the average OP is significantly increasing for four schemes. This is expected since more traveling velocity $V_{\max}$ is needed to compensate the latency requirement which is in contradiction with the $V_{\max}$ limitation.

Next, we examine the energy consumption of the proposed optimization in Section~\ref{Subsec:B} and compare with the TSP-based reference scheme in \cite{Y_Zeng,Y_Zeng_1}. For a fair comparison, we assume that $V_{\max}$ is sufficiently large so that all schemes have at least one path satisfying the RT constraint. Once a feasible set of paths is obtained based on the TSP solution, proposed Algorithms 1, 2 and 3, we apply the optimization $\mathcal{P}_2$ to minimize the total UAV's energy consumption. 

Fig. \ref{fig:6} illustrates the average OP versus energy budget (Joules), where the RT requirements $\eta_k$ ranging between 3 and 15 seconds, $K=4$, $\epsilon=$ 0.001. It is the same with Figs. \ref{fig:2} and \ref{fig:3}, the outage probability value of DP converges to that of the exhaustive search which outperforms the heuristic and reference methods. Furthermore, when the value of energy budget is large enough, the OP of all algorithms converges to the saturation value. It is because the OP is dependent on the energy budget and the RT constraint, as well as the $\mathbb{P}\{R_k < \overline{R}_k\}=\epsilon$, i.e., constraints C1 and C3 in \eqref{OP:energy}. %More specifically, the reference consumes a higher energy in the latency constraints scenario. 

Fig.~\ref{fig:7} plots the energy consumption (Joules) of all schemes as a function of the number of GUs, i.e., $K$, with $\eta_{\min }=3$ seconds, $\eta_{\max }=15$ seconds, $\Psi=K$, $Q_k=10$ Mbits, $\epsilon=0.001,$ $G=15$ dB. A similar observation is that our proposed designs significantly reduce the UAV's consumed energy compared with the reference. This is due to the fact that the reference (TSP-based) always selects the shortest path regardless of the GUs' RT requirements. Consequently, in order to satisfy all GUs' RT constraint, the UAV (in this case) has to fly with a higher velocity than in our proposed designs. Obviously, serving more GUs requires more energy consumption, as shown in these figures. Fig.~\ref{fig:8} describes the details of the maximum and minimum energy consumption for each algorithm.

Fig. \ref{fig:9} evaluates the average energy consumption versus network size with $\eta_{\min }=15$ seconds, $\eta_{\max}=60$ seconds, $\Psi=K=4$, $Q_k=$ 10 Mbits. In Fig. \ref{fig:7}, we assume that the velocity $V_{\max}$ is sufficiently large to make sure that the TSP scheme exists one feasible path which is infeasible in practical. Thus, in Fig. \ref{fig:9}, we compare the average energy consumption of the exhaustive search, heuristic, and DP algorithms with $V_{\max}=50$ m/s. We can observe that, for a larger network size, the energy consumption is increasing. Due to the fact that, the energy consumption depends not only on velocity but also on traveling distance (from eq. \eqref{eq:6}). Specifically, the energy consumption value of DP is very close to that of the exhaustive algorithm. Moreover, the heuristic consumes more energy compared to that of exhaustive search and DP schemes and the gap between them increases proportionally with the network size. 

Last, to illustrate the complexity of all algorithms, Fig. \ref{fig:10} shows the average running time (seconds) as a function of the number of GUs. Clearly, the exhaustive search (Algorithm 1) imposes the largest running time, which increases exponentially with the number of GUs, as it tries all possible paths. The heuristic search (Algorithm 2) and dynamic programing (Algorithm 3) consume much less time compared with Algorithm 1. From practical aspects, Algorithm 3 is preferred as it has a relatively small complexity while achieving good performance. Algorithm 2 consumes less time than Algorithm 3, but it has lower performance, i.e., the OP and energy consumption. Although having a short running time, the TSP-based reference has a poor performance, which is far worse than the proposed Algorithms, as shown in Figs. \ref{fig:3} to \ref{fig:8}.

%%%%%%%%%%%%%%%%%%%%%%%%%%%%%%%%%%%%%%%%%%%%%%%%%%%%%%%%%%%%%%%%%%%%%%
%                              Conclusions                           % 
%%%%%%%%%%%%%%%%%%%%%%%%%%%%%%%%%%%%%%%%%%%%%%%%%%%%%%%%%%%%%%%%%%%%%%
\section{Discussion and Conclusion}
\label{Sec:Con}
\subsection{Discussion}
This research makes a first attempt to design the coarse trajectory for the energy minimization in UAV-enabled wireless communications with latency constraints. The proposed approach can be extended to a fine trajectory, e.g., waypoints based on VBS placement and convex optimization (WVC) \cite{Y_Zeng} and fly-hover-communication (FHC) \cite{Y_Zeng_1}. However, it leads to new challenges since \cite{Y_Zeng} and \cite{Y_Zeng_1} do not take the latency constraints into consideration. Thus, FHC and WVC can not be directly applied to the problem investigated in this paper. Fortunately, the proposed algorithms in our work, i.e., exhaustive, heuristic, and DP in Section III-A, become initial feasible paths for the block coordinate descend (BCD) in combination with the successive convex approximation (SCA) method \cite{Zhan} that can be considered as a new method to obtain the fine trajectory, e.g., FHC and WVC. Moreover, a variable velocity can also be achieved by applying this new approach.
\subsection{Conclusion}
We have investigated the energy-efficient trajectory design for UAV-assisted communications networks which take into consideration latency requirements from the GUs. Concretely, we minimize the total energy consumption via jointly optimizing the UAV trajectory and velocity while satisfying the RT constraints and energy budget. The problem was non-convex, which was solved via two consecutive steps. Firstly, we proposed two algorithms for UAV trajectory design while satisfying the GUs' latency constraints based on the TSPTW. Secondly, for given feasible trajectories, we minimized the total energy consumption via a joint design of the UAV's velocities in all hops. Then, the best path was selected as the designed trajectory of UAV. It was shown via numerical results that our proposed designs outperform the TSP scheme in terms of both energy consumption and outage probability.

The outcome of this work motivates future works in UAV communications networks. One problem is to jointly select the paths and optimize the velocity, which requires advanced optimization techniques but might further improve the UAV's performance. Another promising problem is to consider dynamic network topology. In this case, an adaptive solution that optimizes the UAV trajectory on the fly is required. Furthermore, this research result motivates the trajectory design for multi-UAV scenario, in which multiple UAVs jointly serve the ground users. Pursuing the optimal solution in this case requires advanced optimization techniques and may need collaboration among the UAVs.

%%%%%%%%%%%%%%%%%%%%%%%%%%%%%%%%%%%%%%%%%%%%%%%%%%%%%%%%%%%%%%%%%%%%%%

\section{Acknowledgement}
\label{ACK}

This research is supported by the Luxembourg National Research Fund under project FNR CORE ProCAST, grant C17/IS/11691338 and FNR CORE 5G-Sky, grant C19/IS/13713801.

%                              References                            % 
%%%%%%%%%%%%%%%%%%%%%%%%%%%%%%%%%%%%%%%%%%%%%%%%%%%%%%%%%%%%%%%%%%%%%%% ---- Bibliography ----

%\bibliographystyle{IEEEtran}
%\bibliography{IEEEabrv,bibJournalList,conf_cite}

\begin{IEEEbiography}[{\includegraphics[width=1in,height=1.25in,clip,keepaspectratio]{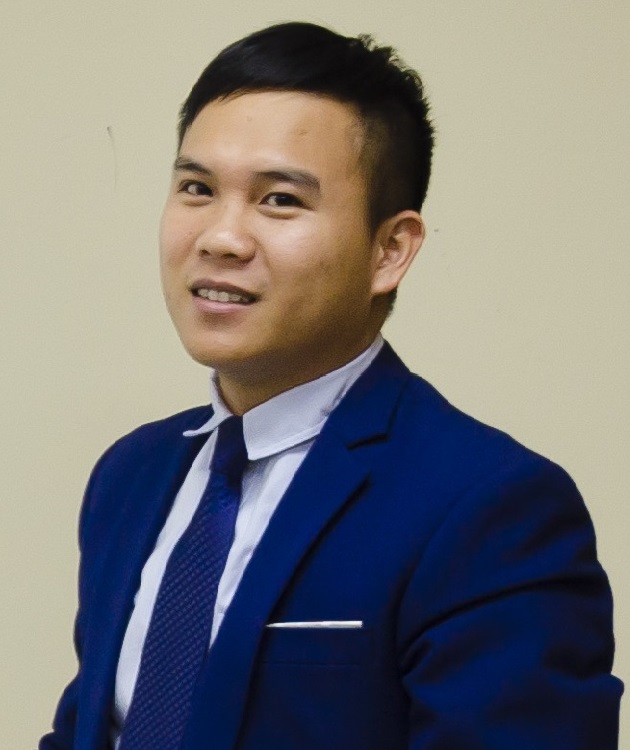}}]
	{Dinh-Hieu Tran},  (S'20)  was born and grew up in Gia Lai, Vietnam (1989). He received the B.E. degree in Electronics and Telecommunication Engineering Department from Ho Chi Minh City University of Technology, Vietnam, in 2012. In 2017, he finished the M.Sc degree in Electronics and Computer Engineering from Hongik University (Hons.), South Korea. He is currently pursuing the Ph.D. degree at the Interdisciplinary Centre for Security, Reliability and Trust (SnT), University of Luxembourg, under the supervision of Prof. Symeon Chatzinotas and Prof. Bj\"orn Ottersten. His research interests include UAVs, IoTs, Backscatter, B5G for wireless communication networks.  He was a recipient of the IS3C 2016 best paper award.
\end{IEEEbiography}
\begin{IEEEbiography}[{\includegraphics[width=1in,height=1.25in,clip,keepaspectratio]{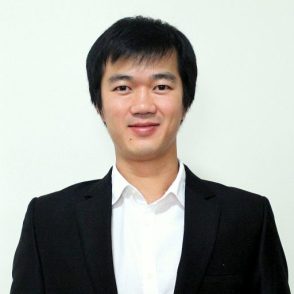}}]
{Thang X. Vu} (M'15) was born in Hai Duong, Vietnam. He received the B.S. and the M.Sc., both in Electronics and Telecommunications Engineering, from the VNU University of Engineering and Technology, Vietnam, in 2007 and 2009, respectively, and the Ph.D. in Electrical Engineering from the University Paris-Sud, France, in 2014. 

In 2010, he received the Allocation de Recherche fellowship to study Ph.D. in France. From September 2010 to May 2014, he was with the Laboratory of Signals and Systems (LSS), a joint laboratory of CNRS, CentraleSupelec and University Paris-Sud XI, France. From July 2014 to January 2016, he was a postdoctoral researcher with the Information Systems Technology and Design (ISTD) pillar, Singapore University of Technology and Design (SUTD), Singapore. Currently, he is a research scientist at the Interdisciplinary Centre for Security, Reliability and Trust (SnT), University of Luxembourg, Luxembourg. His research interests are in the field of wireless communications, with particular interests of 5G networks and beyond, machine learning for communications and cross-layer resources optimization. He was a recipient of the SigTelCom 2019 best paper award.
\end{IEEEbiography}

\begin{IEEEbiography}[{\includegraphics[width=1in,height=1.25in,clip,keepaspectratio]{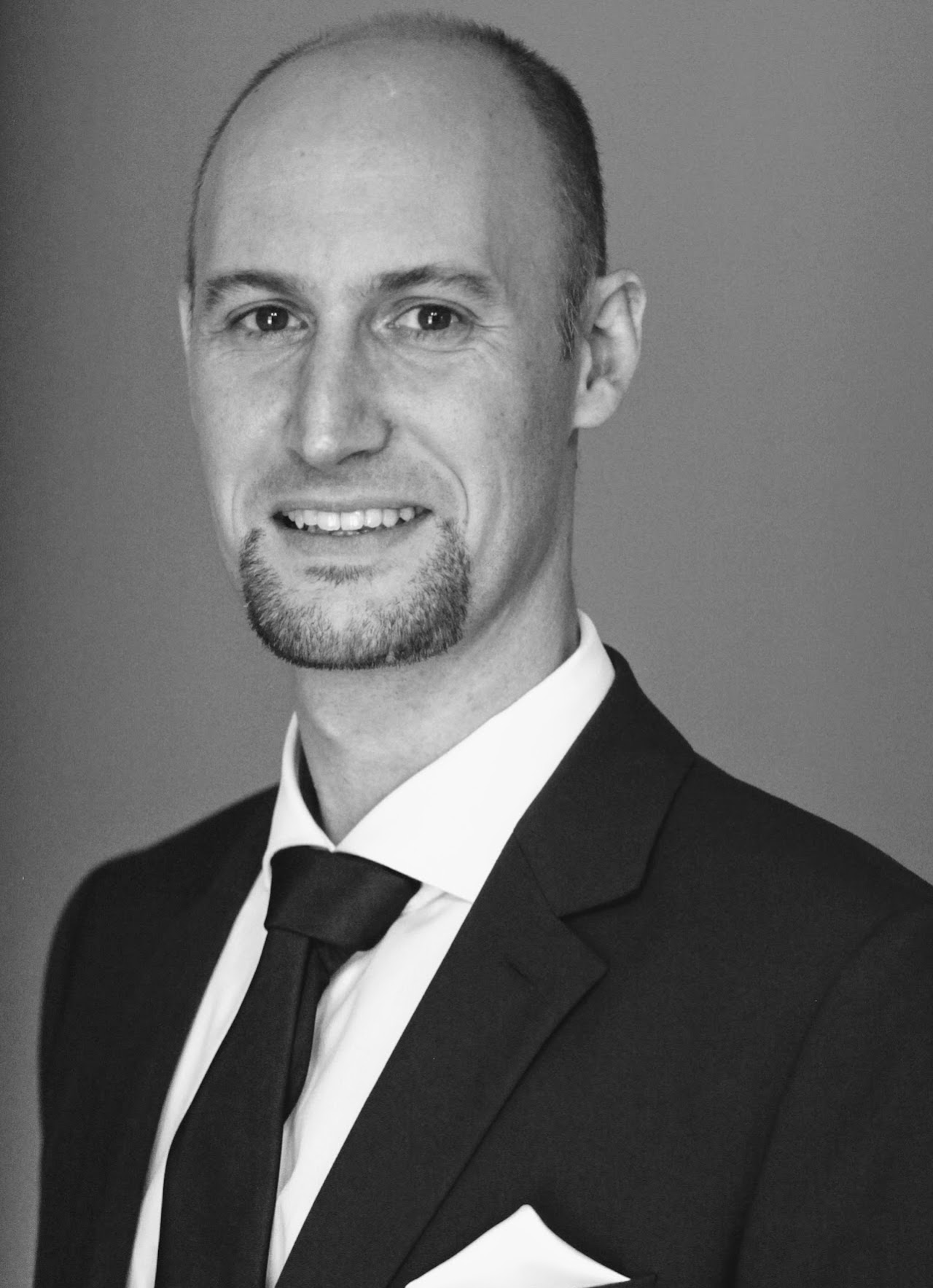}}]
	{Symeon Chatzinotas}, (S'06-M'09-SM'13) is currently Full Professor / Chief Scientist I and Co-Head of the SIGCOM Research Group at SnT, University of Luxembourg. In the past, he has been a Visiting Professor at the University of Parma, Italy and he was involved in numerous Research and Development projects for the National Center for Scientific Research Demokritos, the Center of Research and Technology Hellas and the Center of Communication Systems Research, University of Surrey. He received the M.Eng. degree in telecommunications from the Aristotle University of Thessaloniki, Thessaloniki, Greece, in 2003, and the M.Sc. and Ph.D. degrees in electronic engineering from the University of Surrey, Surrey, U.K., in 2006 and 2009, respectively. He was a co-recipient of the 2014 IEEE Distinguished Contributions to Satellite Communications Award,  the CROWNCOM 2015 Best Paper Award and the 2018 EURASIC JWCN Best Paper Award. He has (co-)authored more than 400 technical papers in refereed international journals, conferences and scientific books. He is currently in the editorial board of the IEEE Open Journal of Vehicular Technology and the International Journal of Satellite Communications and Networking.
\end{IEEEbiography}	

\begin{IEEEbiography}[{\includegraphics[width=1in,height=1.25in,clip,keepaspectratio]{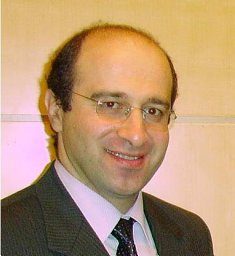}}]
	{Shahram ShahbazPanahi}, (M'02-SM'10) was born in Sanandaj, Kurdistan, Iran. He received the B.Sc., M.Sc., and Ph.D. degrees in electrical engineering from Sharif University of Technology, Tehran, Iran, in 1992, 1994, and 2001, respectively. From September 1994 to September 1996, he was an instructor with the Department of Electrical Engineering, Razi University, Kermanshah, Iran. From July 2001 to March 2003, he was a Postdoctoral Fellow with the Department of Electrical and Computer Engineering, McMaster University, Hamilton, ON, Canada. From April 2003 to September 2004, he was a Visiting Researcher with the Department of Communication Systems, University of Duisburg-Essen, Duisburg, Germany. From September 2004 to April 2005, he was a Lecturer and Adjunct Professor with the Department of Electrical and Computer Engineering, McMaster University. In July 2005, he joined the Faculty of Engineering and Applied Science, University of Ontario Institute of Technology, Oshawa, ON, Canada, where he currently holds a Professor position. His research interests include statistical and array signal processing; space-time adaptive processing; detection and estimation; multi-antenna, multi-user, and cooperative communications; spread spectrum techniques; DSP programming; and hardware/realtime software design for telecommunication systems. Dr. Shahbazpanahi has served as an Associate Editor for the IEEE TRANSACTIONS ON SIGNAL PROCESSING and the IEEE SIGNAL PROCESSING LETTERS. He has also served as a Senior Area Editor for the IEEE SIGNAL PROCESSING LETTERS. He was an elected member of the Sensor Array and Multichannel (SAM) Technical Committee of the IEEE Signal Processing Society. He has received several awards, including the Early Researcher Award from Ontario’s Ministry of Research and Innovation, the NSERC Discovery Grant (three awards), the Research Excellence Award from the Faculty of Engineering and Applied Science, the University of Ontario Institute of Technology, and the Research Excellence Award, Early Stage, from the University of Ontario Institute of Technology.
\end{IEEEbiography}		
	
\begin{IEEEbiography}[{\includegraphics[width=1in,height=1.25in,clip,keepaspectratio]{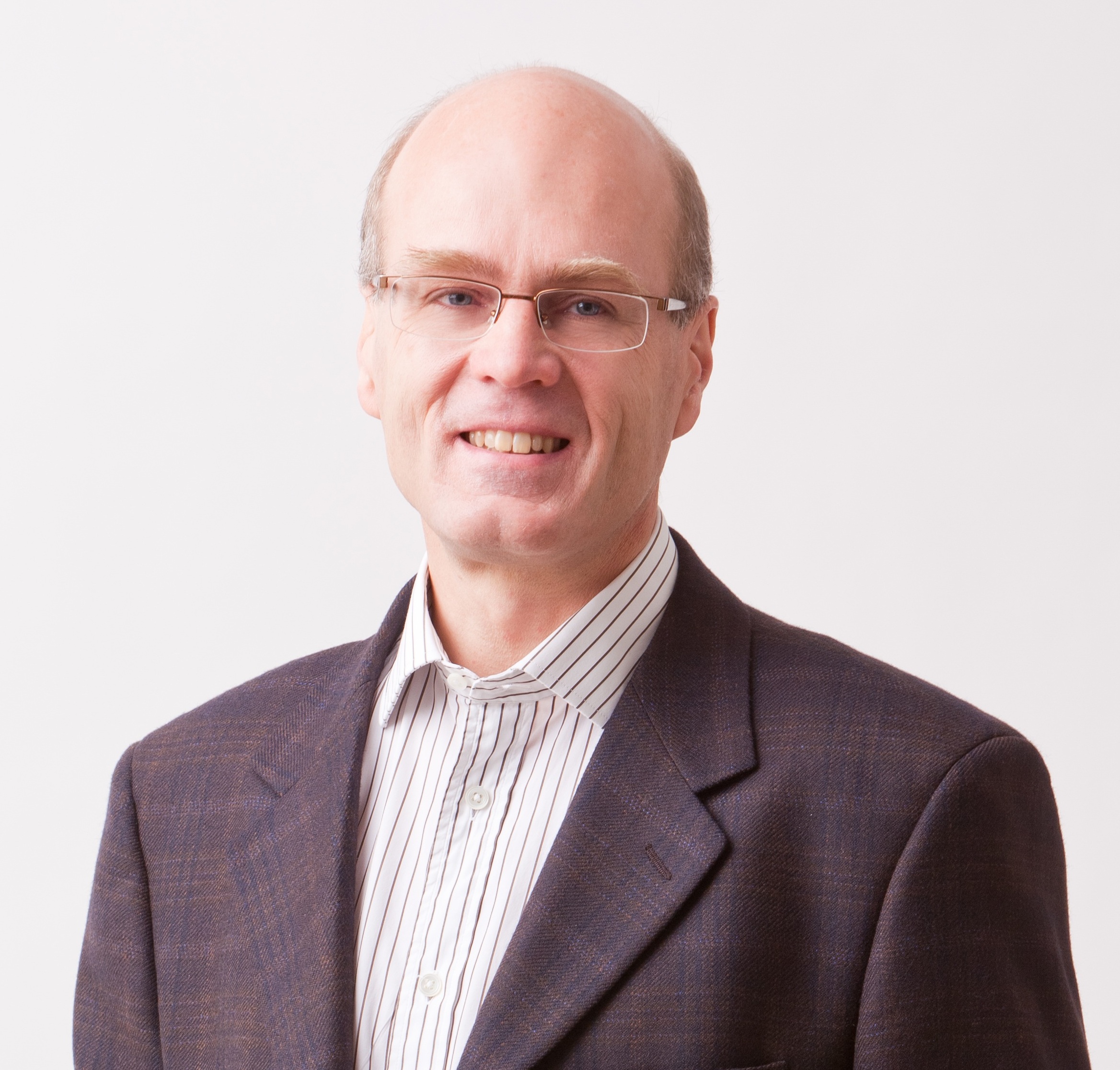}}]
	{Bj\"orn Ottersten}, (S'87-M'89-SM'99-F'04) was born in Stockholm, Sweden, in 1961. He received
	the M.S. degree in electrical engineering and applied physics from Linköping University, Linköping, Sweden, in 1986, and the Ph.D. degree in electrical engineering from Stanford University, Stanford, CA, USA, in 1990. He has held research positions with the Department of Electrical Engineering, Linköping
	University, the Information Systems Laboratory, Stanford University, the Katholieke Universiteit	Leuven, Leuven, Belgium, and the University of Luxembourg, Luxembourg. From 1996 to 1997, he was the Director of Research with ArrayComm, Inc., a start-up in San Jose, CA, USA, based on his patented technology. In 1991, he was appointed Professor of signal processing with the Royal Institute of Technology (KTH), Stockholm, Sweden. Dr. Ottersten has been Head of the Department for Signals, Sensors, and Systems, KTH, and Dean of the School of Electrical Engineering, KTH. He is currently the Director for the Interdisciplinary Centre for Security, Reliability and Trust, University of Luxembourg.
	He is a recipient of the IEEE Signal Processing Society Technical Achievement Award and the European Research Council advanced research grant twice. He has co-authored journal papers that received the IEEE Signal Processing Society Best Paper Award in 1993, 2001, 2006, 2013, and 2019, and 8 IEEE conference papers best paper awards. He has been a board member of IEEE Signal Processing Society, the Swedish Research Council and currently serves of the boards of EURASIP and the Swedish Foundation for Strategic Research. He has served as an Associate Editor for the IEEE TRANSACTIONS ON SIGNAL PROCESSING and the Editorial Board of the IEEE Signal Processing Magazine. He is currently a member of the editorial boards of IEEE Open Journal of Signal Processing, EURASIP Signal Processing Journal, EURASIP Journal of Advances Signal Processing and Foundations and Trends of Signal Processing. He is a fellow of EURASIP.	
\end{IEEEbiography}

\end{document}